\documentclass
[a4paper,superscriptaddress,twocolumn,prb,preprintnumbers,floatfix,secnumarabic,amssymb]{revtex4-1}
\usepackage{todonotes}

\usepackage{subfigure}
\usepackage[english]{babel}
\usepackage{caption}
\usepackage[utf8]{inputenc}
\usepackage{dsfont}
\usepackage{mathtools}
\usepackage{amssymb}
\usepackage{amsthm}
\usepackage{mathrsfs}
\usepackage{mathdots}
\newcommand{\e}[1]{\mathrm{e}^{#1}}
\usepackage{ifthen}
\usepackage{cancel}
\usepackage{color}
\usepackage[colorlinks, linkcolor = blue, citecolor = blue, filecolor = black, urlcolor = blue]{hyperref}
\usepackage[super]{nth}
\usepackage{grffile} 
\usepackage{floatflt} 
\usepackage[export]{adjustbox} 
\usepackage[]{natbib}

\usepackage[]{units}
\usepackage{xspace}  

\newcommand \newk {k}

\newcommand \myEquBegin{\begin{equation}\begin{aligned}[b]}
\newcommand \myEquEnd{\end{aligned}\end{equation}}
\newcounter{QMD}
\newcommand \QMDraw{\ifthenelse{\equal{\arabic{QMD}}{0}}{quasi momentum distribution (QMD)\setcounter{QMD}{1}}{QMD}}
\newcommand \QMD{\QMDraw\xspace}

\newcounter{QSQMD}
\newcommand \QSQMD{\ifthenelse{\equal{\arabic{QSQMD}}{0}}{``quasi-stationary QMD'' (QSQMD)\xspace\setcounter{QSQMD}{1}}{QSQMD\xspace}}
\newcounter{FHM}
\newcommand \FHM{\ifthenelse{\equal{\arabic{FHM}}{0}}{Fermi-Hubbard model (FHM)\xspace\setcounter{FHM}{1}}{FHM\xspace}}
\newcounter{NNNHterm}
\newcommand \NNNHterm{\ifthenelse{\equal{\arabic{NNNHterm}}{0}}{next-to-nearest-neighbor-hopping-term (NNNH)\xspace\setcounter{NNNHterm}{1}}{NNNH\xspace}}
\newcounter{BE}
\newcommand \BE{\ifthenelse{\equal{\arabic{BE}}{0}}{quantum Boltzmann equation (BE)\xspace\setcounter{BE}{1}}{BE\xspace}}
\newcommand \BoltzmannEquation {Boltzmann equation\xspace}
\newcounter{LBE}
\newcommand \LBE{\ifthenelse{\equal{\arabic{LBE}}{0}}{linearized Boltzmann-equation (LBE)\xspace\setcounter{LBE}{1}}{LBE\xspace}}
\newcommand \smns{\!-\!}
\newcommand \spls{\!+\!}
\newcommand \kB {k_B}
\DeclareMathOperator \const{const}

\newcommand \vphdagger{{\vphantom{\dagger}}}
\newcommand \whitedagger{{\color{white} \dagger}}

\definecolor{marker}{rgb}{.7,1,1}

\newcommand \fdnude{f}

\newcommand \fd {\fdnude}

\newcommand \emu[1] {e^{\ifthenelse{\equal{#1}{+}}{}{#1} \beta \mu}}

\newcommand \n {\hat{n}}
\newcommand \N {\hat{N}}
\newcommand \XYvec[2] {{ \vec{#1} , \vec{#2} }}
\newcommand \nkvec {{ \XYvec{\nb}{k} }}

\newcommand \updownXY[4][NONE]{{{%
	\ifthenelse{\equal{#1}{NONE}}{%
		#2 #3 #4%
	}{%
		#2 #3_{#1} #4_{#1}%
	}%
}}}

\newcommand \phink[1]{ \fctnk{\phi}{#1} }

\newcommand \dphinktfct[1]{ \indfct{\dot\phi}{\nb}{k,t} }
\newcommand \Domega {W} 

\newcommand \Domegank {\Domega_{\!\nkvec}}

\newcommand \dK {P} 

\newcommand \dKX[1][\kvec]{\dK_{#1}}
\newcommand \ddKX[1][\kvec]{\delta(\dKX[#1])}

\newcommand \dkX[2][k] {
	\ifthenelse{#2>3}{d^{#2}\hspace{-1.5pt}#1}{
	\ifthenelse{#2>1}{d#1_2}{} 
	\ifthenelse{#2>0}{d#1_3}{d#1}
	\ifthenelse{#2>2}{d#1_4}{}
	}
}

\newcommand \order {\mathcal{O}}

\newcommand \fscalprod[2]{\langle #1, #2 \rangle_{\findex}}
\newcommand \fnorm[1]{\| #1 \|_{\findex}}
\renewcommand \fscalprod[2]{\langle #1, #2 \rangle}
\renewcommand \fnorm[1]{\| #1 \|}
\newcommand \Icoll { \mathcal{I}_{\text{coll}}} 
\newcommand \Lc{\LcRAW}
\newcommand \LcRAW{\hat{\mathcal{L}}}
\newcommand \nb{\nu} 
\newcommand \fct[3][]{#2#1(#3)}
\newcommand \indfct[4][\hspace{-1pt}]{\fct[#1]{#2_{#3}}{#4}}
\newcommand \fctnk[2]{
	\fctsnk{#1}{}{#2}
}
\newcommand \ifEqualOneTillFour[3]{
	\ifthenelse{\equal{#1}{1} \or \equal{#1}{2} \or \equal{#1}{3} \or \equal{#1}{4}}{#2}{#3}
}
\newcommand \fctsk[3]{
	\ifthenelse{\equal{#3}{0}}{
		\indfct{#1}{#2}{k}
	}{
		\ifEqualOneTillFour{#3}{
			\indfct{#1}{#2}{k_{#3}}
		}{
			\indfct{#1}{#2}{#3}
		}
	}
}
\newcommand \fctsnk[3]{
	\ifthenelse{\equal{#3}{0}}{
		\indfct{#1}{#2\nb}{k}
	}{
		\indfct{#1}{#2\nb_{#3}}{k_{#3}}
	}
}
\newcommand \intlimdk[2][k] {
	\int
	\ifthenelse{#2>1}{
		\limits_{\mathclap{\qquad [-1/2,1/2]^{#2}}}
	}{
		\limits_{\mathclap{-1/2}}^{\mathclap{1/2}}
		\!
	}
	\dk[#1]{#2}
}
\newcommand \intlimdkS[2][k] {
	\int
	\ifthenelse{#2>1}{
		\limits_{\mathclap{\IBZ^{#2}}}
	}{
		\limits_{\mathclap{-1/2}}^{\mathclap{1/2}}
		\!
	} \!
	\dkS[#1]{#2}
}
\newcommand \intlimdkX[2][k] {
	\int
	\ifthenelse{#2>1}{
		\limits_{\mathclap{\IBZ^{#2}}}
		\!
	}{
		\limits_{\mathclap{-1/2}}^{\mathclap{1/2}}
		\!
	}
	\dkX[#1]{#2}
}
\newcounter{intlimdkZ}
\newcommand \intlimBZ[1] {
	\int
	\ifthenelse{#1>1}{
		\ifthenelse{#1>2}{
			\setcounter{intlimdkZ}{#1}
			\addtocounter{intlimdkZ}{-1}
			\limits_{\mathclap{\IBZ^{\arabic{intlimdkZ}}\times \mathds R}}
			\!
		}{
			\limits_{\mathclap{\IBZ^{#1}}}
			\!
		}
	}{
		\limits_{\mathclap{-1/2}}^{\mathclap{1/2}}
		\!
	}
}

\newcommand \ppmm{\mathbin{\vcenter{\hbox{%
  \oalign{$\scriptscriptstyle+$\cr
          \noalign{\kern-0ex}
          $\scriptscriptstyle{+}$\cr%
          \noalign{\kern-.3ex}
          $\scriptscriptstyle{-}$\cr%
          \noalign{\kern-.6ex}
          $\scriptscriptstyle{-}$\cr}%
}}}}
\newcommand \pmmp{\mathbin{\vcenter{\hbox{%
  \oalign{$\scriptscriptstyle+$\cr%
          \noalign{\kern-.3ex}
          $\scriptscriptstyle{-}$\cr%
          \noalign{\kern-.6ex}
          $\scriptscriptstyle{-}$\cr
          \noalign{\kern-.3ex}
          $\scriptscriptstyle{+}$\cr}%
}}}}
\newcommand \mppm{\mathbin{\vcenter{\hbox{%
  \oalign{$\scriptscriptstyle-$\cr%
          \noalign{\kern-.3ex}
          $\scriptscriptstyle{+}$\cr%
          \noalign{\kern-0ex}
          $\scriptscriptstyle{+}$\cr
          \noalign{\kern-.3ex}
          $\scriptscriptstyle{-}$\cr}%
}}}}

\newcommand \pmpm{\mathbin{\vcenter{\hbox{%
  \oalign{$\scriptscriptstyle+$\cr
          \noalign{\kern-.3ex}
          $\scriptscriptstyle{-}$\cr%
          \noalign{\kern-.3ex}
          $\scriptscriptstyle{+}$\cr%
          \noalign{\kern-.3ex}
          $\scriptscriptstyle{-}$\cr}%
}}}}

\newcommand\mpmp{\mathbin{\vcenter{\hbox{%
  \oalign{$\scriptscriptstyle-$\cr
          \noalign{\kern-.3ex}
          $\scriptscriptstyle{+}$\cr%
          \noalign{\kern-.3ex}
          $\scriptscriptstyle{-}$\cr%
          \noalign{\kern-.3ex}
          $\scriptscriptstyle{+}$\cr}%
}}}}

\newcommand \up {\uparrow}
\newcommand \down {\downarrow}

\newcommand \kvec {{\vec{k}}}

\newcommand \nbvec {{\vec{\nb}}}
\newcommand \Sb{\mathds{B}} 
\newcommand \IBZ{\mathds{K}} 

\newcommand \hopping {t_{hop}}

\newcommand \lamX {\lambda}

\newcommand \DelX{\Delta}
\newcommand \upchi{\text{\protect\raisebox{1.8pt}{$\chi$}}}
\newcommand \EFraw{\upchi}

\newcommand \EFj[1]{\EFraw^{(#1)}}

\newcommand \indfctjnk[3]{{
	\ifthenelse{\equal{#3}{0}}{
		\indfct{#2^{(#1)}}{\nb}{k}
	}{
		\indfct{#2^{(#1)}}{\nb_{#3}}{k_{#3}}
	}
}}
\newcommand \psink[1]{\fctnk{\psi}{#1}}
\newcommand \EFjnk[2]{\indfctjnk{#1}{\EFraw}{#2}}

\newcommand \Phiint {\Phi}

\newcommand \Ham {\hat{H}}

\newcommand \HO { \Ham_0}
\newcommand \Hint {\Ham_{\text{int}}}
\newcommand \cRaw {\hat{c}}
\newcommand \cd {\cRaw^{\dagger}}
\newcommand \cw {\cRaw^{\vphdagger}}
\newcommand \SetEpsToTwoJNNNOverJNN {0}
\ifthenelse{\equal{\SetEpsToTwoJNNNOverJNN}{1}}{

}{

}
\newcommand \fctnkX[4][]{
	\ifthenelse{\equal{#3}{0}}{
		\ifthenelse{\equal{#4}{0}}{
			\indfct{#2}{\nb}{k #1}
		}{
			\indfct{#2}{\nb}{k_{#4} #1}
		}
	}{
		\ifthenelse{\equal{#4}{0}}{
			\indfct{#2}{\nb_{#3}}{k #1}
		}{
			\indfct{#2}{\nb_{#3}}{k_{#4} #1}
		}
	}
}
\newcommand \fctnkt[3][t]{
	\ifthenelse{\equal{#3}{0}}{
		\indfct{#2}{\nb}{k,#1}
	}{
		\indfct{#2}{\nb_{#3}}{k_{#3},#1}
	}
}
\newcommand \fctsnkt[3][t]{
	\ifthenelse{\equal{#3}{0}}{
		\indfct{#2}{\sn}{k,#1}
	}{
		\indfct{#2}{\sn_{#3}}{k_{#3},#1}
	}
}

\newcommand \phinkt[2][t]{ \fctnkt[#1]{\phi}{#2} }
\newcommand \dotphinkt[2][t]{ \fctnkt[#1]{\dot\phi}{#2} }
\newcommand \Phiintnk{\Phiint_{\nbvec,\kvec}}
\newcommand \PhiintnkII{\bigl| \Phiintnk \bigr|^2}

\newcommand \sn{{\sigma \nb}}

\newcommand \fdnk[1]{ \fctnk{\fd}{#1} }

\newcommand \ARaw{A}
\newcommand \AjO[1]{\ARaw_{#1}(0)}
\newcommand\hPhi{\hat{\Phi}}

\newcommand\SSop{\hPhi}  

\newcommand \putFigureRelaxationRates[1]{
\begin{figure*}
\begin{center}
\newcommand \FigRelaxRates {133pt}
\includegraphics[height=\FigRelaxRates,trim=0 0 95 0,clip]
	{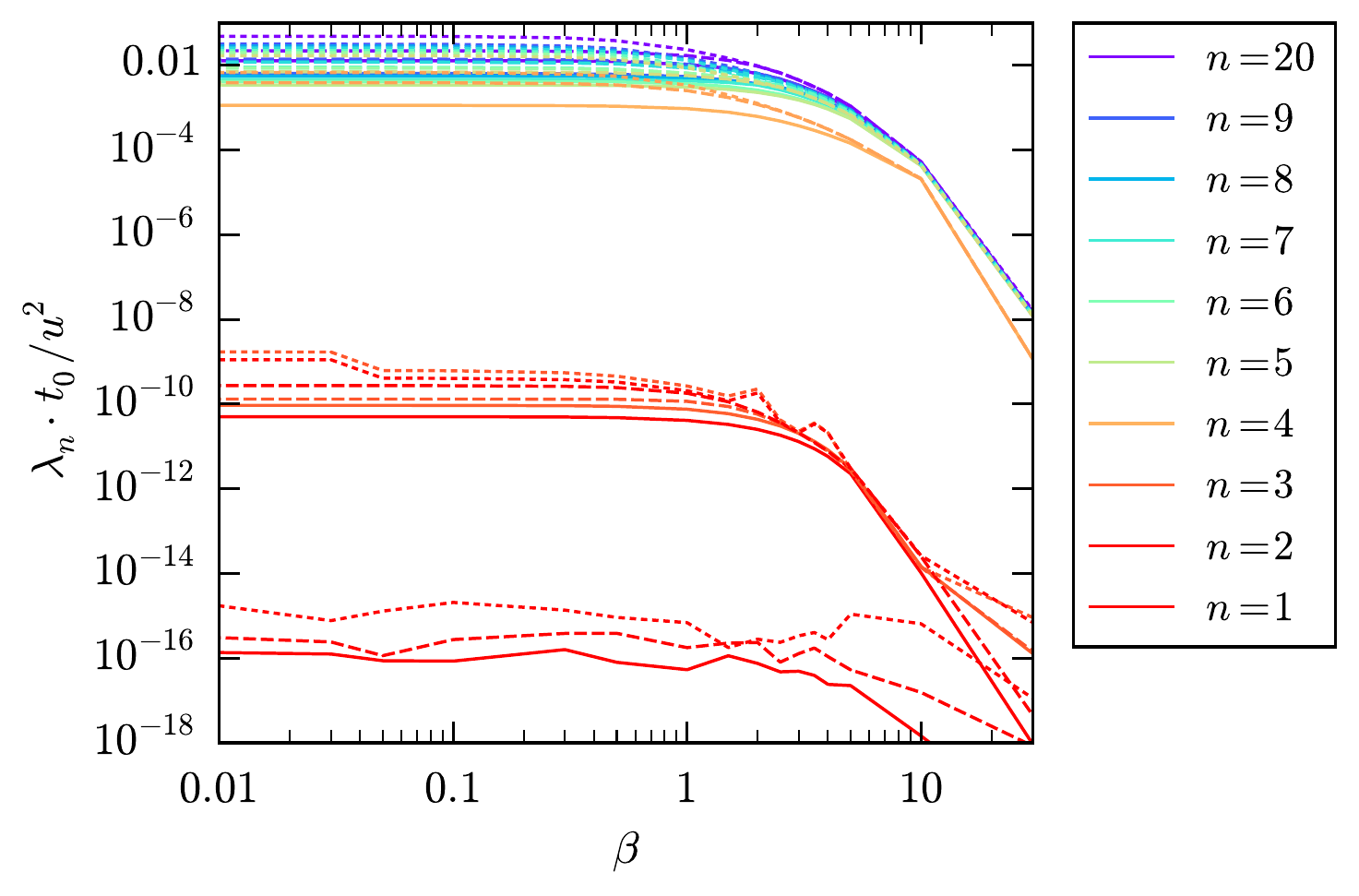}
\includegraphics[height=\FigRelaxRates,trim={25 0 95 0},clip]
	{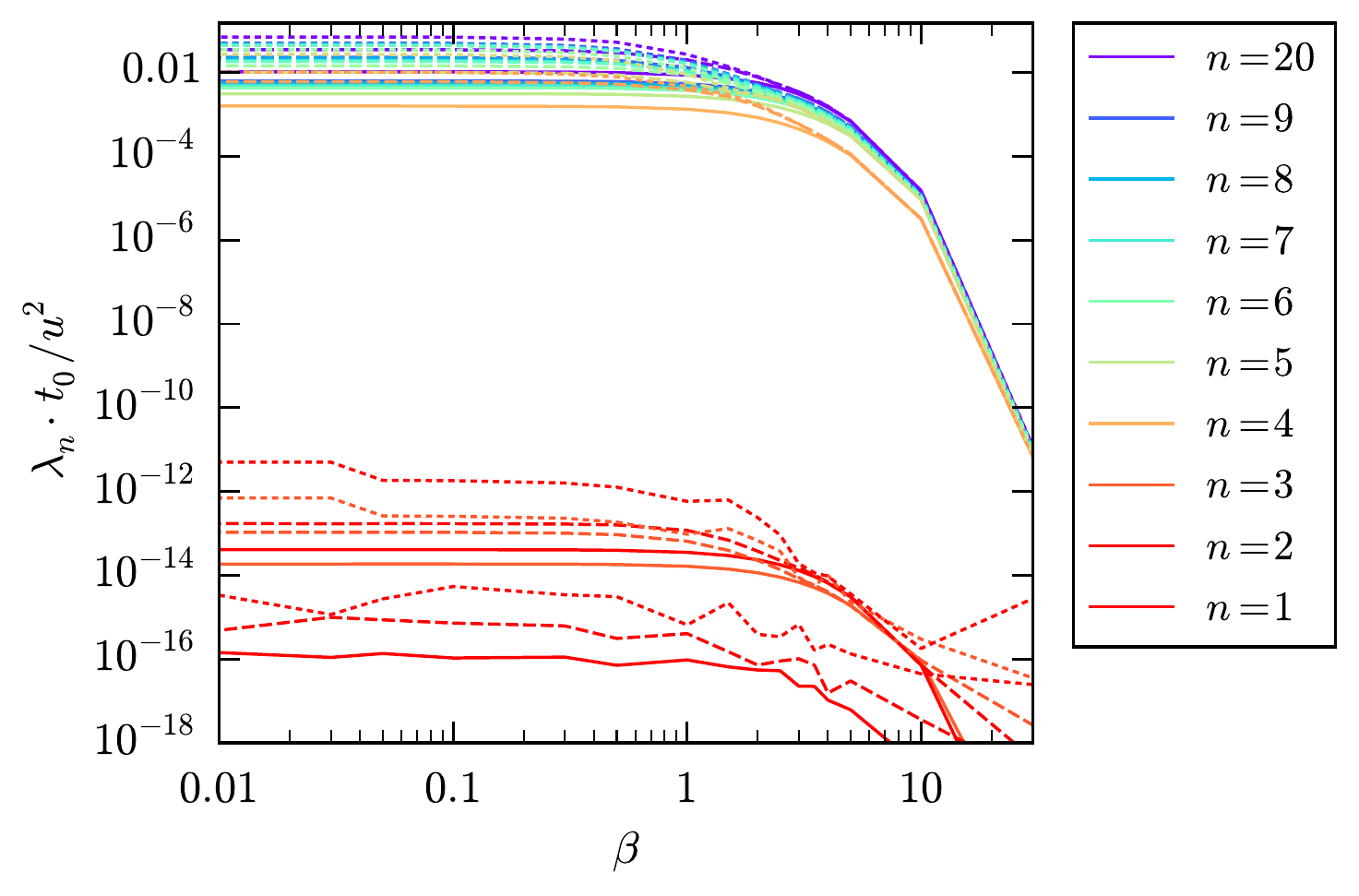}
\includegraphics[height=\FigRelaxRates,trim={25 0 0 0},clip]
	{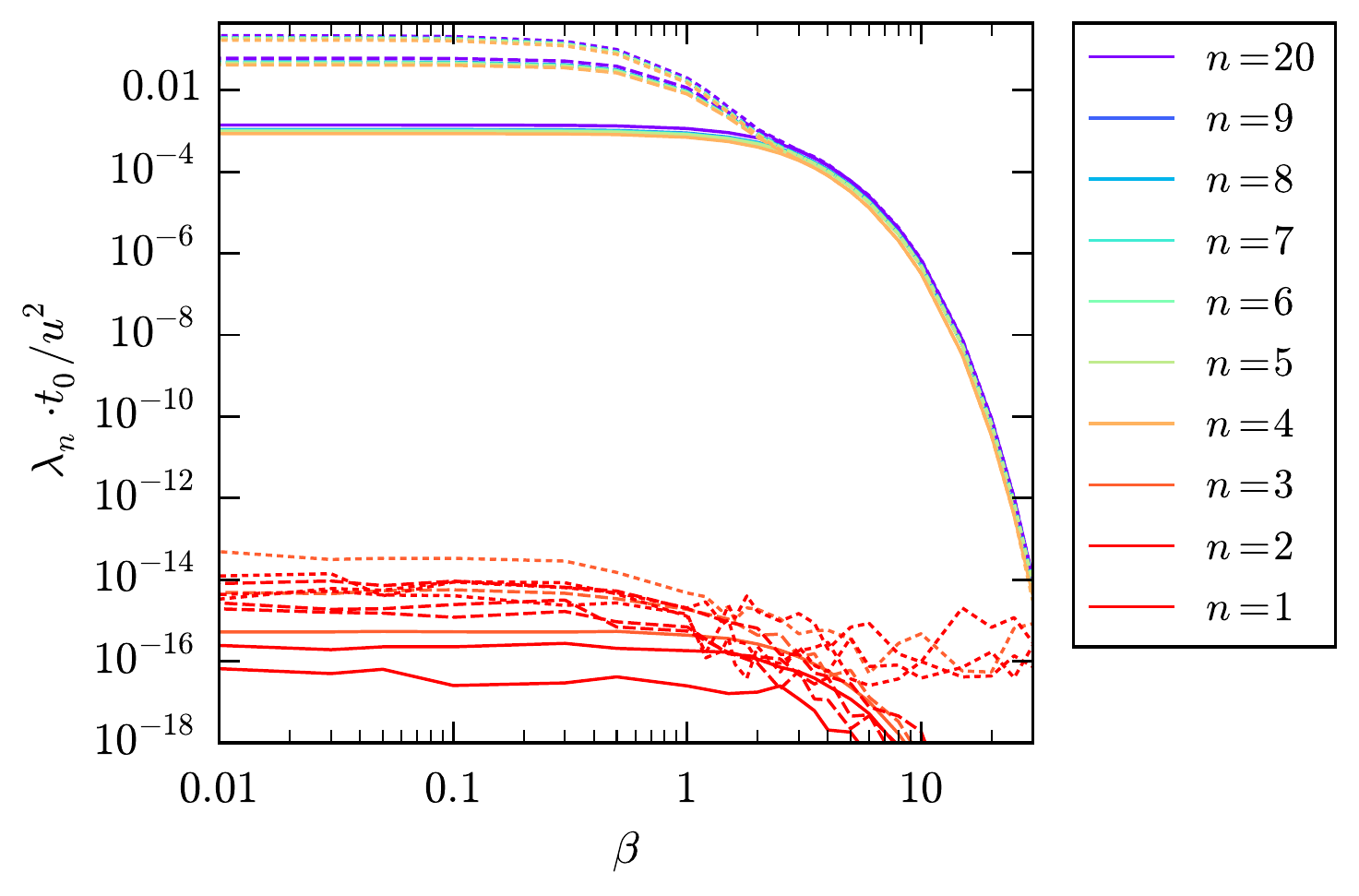}
\caption{#1}
\end{center}
\end{figure*}
}

\newcommand \betaRelaxTimeSplitting{1}

\newcommand \nel {n^{\text{el}}}

\newcommand \nqp {n^{\text{qp}}}

\newcommand \dotnqp {\dot{n}^{\text{qp}}}

\usepackage{ulem}
\usepackage{nicefrac}

\usepackage{bm}

\usepackage{cleveref}  
\Crefname{appendix}{Appendix}{Appendices}
\Crefname{equation}{Equation}{Equations}
\Crefname{figure}{Figure}{Figures}
\Crefname{section}{Section}{Sections}
\Crefname{tabular}{Tabular}{Tabulars}

\crefname{appendix}{Appendix}{Apps.}
\crefname{equation}{Eq.}{Eqs.}
\crefname{figure}{Fig.}{Figs.}
\crefname{section}{Sec.}{Secs.}
\crefname{tabular}{Table}{Tabs.}

\begin{document}

\title{Relaxation of photoexcitations in polaron-induced magnetic microstructures}

\author{Thomas K\"ohler}
\affiliation{Institut f\"ur Theoretische Physik, Universit\"at G\"ottingen, 37077 G\"ottingen, Germany}

\author{Sangeeta Rajpurohit}
\affiliation{Institut f\"ur Theoretische Physik, TU Clausthal, 38678 Clausthal-Zellerfeld, Germany}

\author{Ole Schumann}
\affiliation{Institut f\"ur Theoretische Physik, Universit\"at G\"ottingen, 37077 G\"ottingen, Germany}
\affiliation{Daimler AG, Wilhelm-Runge-Str. 11, 89081 Ulm, Germany}

\author{Sebastian Paeckel}
\affiliation{Institut f\"ur Theoretische Physik, Universit\"at G\"ottingen, 37077 G\"ottingen, Germany}

\author{Fabian R. A. Biebl}
\affiliation{Institut f\"ur Theoretische Physik, Universit\"at G\"ottingen, 37077 G\"ottingen, Germany}
\affiliation{Math2Market, Richard-Wagner-Straße 1, 67655 Kaiserslautern, Germany}

\author{Mohsen Sotoudeh}
\affiliation{Institut f\"ur Theoretische Physik, TU Clausthal, 38678 Clausthal-Zellerfeld, Germany}

\author{Stephan C. Kramer}
\affiliation{Institut f\"ur Theoretische Physik, Universit\"at G\"ottingen, 37077 G\"ottingen, Germany}
\affiliation{Fraunhofer-Institut f\"ur Techno- und Wirtschaftsmathematik ITWM, 67663 Kaiserslautern, Germany}

\author{Peter E. Bl\"ochl}
\affiliation{Institut f\"ur Theoretische Physik, TU Clausthal, 38678 Clausthal-Zellerfeld, Germany}
\affiliation{Institut f\"ur Theoretische Physik, Universit\"at G\"ottingen, 37077 G\"ottingen, Germany}

\author{Stefan Kehrein}
\affiliation{Institut f\"ur Theoretische Physik, Universit\"at G\"ottingen, 37077 G\"ottingen, Germany}

\author{Salvatore R. Manmana}
\affiliation{Institut f\"ur Theoretische Physik, Universit\"at G\"ottingen, 37077 G\"ottingen, Germany}

\date{\today}

\begin{abstract}
We investigate the evolution of a photoexcitation in correlated materials over a wide range of time scales.
The system studied is a one-dimensional model of a manganite with correlated electron, spin, orbital, and lattice degrees of freedom, which we relate to the three-dimensional material Pr$_{1-x}$Ca$_{x}$MnO$_3$. 
The ground-state phases for the entire composition range are determined and rationalized by a coarse-grained polaron model.
At half-doping a pattern of antiferromagnetically coupled Zener polarons is realized. 
Using time-dependent density-matrix renormalization group (tDMRG), we treat the electronic quantum dynamics following the excitation.
The emergence of quasiparticles is addressed, and the relaxation of the nonequilibrium quasiparticle distribution is investigated via a linearized quantum-Boltzmann equation. 
Our approach shows that the magnetic microstructure caused by the Zener polarons leads to an increase of the relaxation times of the excitation.
\end{abstract}


\maketitle

\section{Introduction}

The relaxation of optical excitations in materials is a process that is central for energy conversion in materials. 
In particular, short length and time scales reveal a whole range of interesting physical effects.
In the presence of strong correlations, this topic is one of ongoing experimental\cite{Rini2007,1367-2630-18-9-093028,Schmitt1649,Tao62,Mitrano2016,Hu2014} and theoretical studies.\cite{PhysRevLett.111.016401, PhysRevB.92.201104,Eckstein2013,Eckstein2016} 
A detailed understanding of these processes is expected to open doors for new technological applications.

For example, the controlled application of pump-probe setups on the femtosecond time scale has lead to interesting discoveries, such as the formation of metastable states, some of which are described to be superconducting.\cite{Fausti189} 
Light irradiation of interfaces of correlated materials has shown the possibility to realize unconventional photovoltaic effects\cite{Zhao2005,Zhao2006,Ni2012,Snaith2013, Gong2015,Saucke2012,Ifland2015} that are not based on the formation of excitons, but rather of polarons, i.e., quasiparticles consisting of electrons and phonons that are formed or excited by light absorption.
Pump-probe experiments in manganites have produced evidence for long-lived hot-polaron states.\cite{nanosecondlifetimes}
Such long-lived states are of interest, because they have the potential to overcome the Shockley-Queisser limit\cite{shockley61_jap32_510} for the efficiency of solar cells. 

The relaxation processes of excitations span a wide range of time scales: The absorption process of light in pump-probe experiments can last as short as femtoseconds, whereas the perturbed polaronic order may persist up to time scales in the range of nanoseconds.  
The focus on strong correlations is particularly promising in the context of identifying slow relaxation processes.
Finding a unifying description of the evolution of these excitations, particularly in the presence of strong correlations, is a major challenge. 

In this paper, we combine a number of theoretical approaches to cover the large range of time scales of a relaxation process for a specific material. 
The material chosen has been inspired by manganites, a class of materials with strong correlations between electrons, spins and phonons. 
In order to make the problem tractable, however, we have chosen a one-dimensional model-manganite system, which nevertheless contains many of the relevant properties of the real materials.
We investigate the ground-state properties of this model system and find that they are well rationalized in terms of polaronic order.
Using the time-dependent density-matrix renormalization group (tDMRG) in a matrix product state (MPS) formulation,\cite{white1992,white1993,Schollwock:2005p2117,daley04,white04,Schollwock:2011p2122} we then investigate the time evolution following a dipole excitation, by which we model the effect of a photoexcitation on one of the polaronic ground states. 
This allows us to study the role of the electron-electron interaction in the short-time dynamics after the excitation.
The long-time behavior of the electron relaxation is then investigated using a linearized quantum Boltzmann equation (LBE).\cite{Biebl2016}
The tDMRG and the LBE approaches both show that the relaxation time scales increase with the strength of the magnetic microstructure, which is induced by the polaronic order.

The paper is organized as follows.
A 1D tight-binding model for the model-manganite, its polaronic and magnetic order, and the resulting effective Hubbard-type model are presented in Sec.~\ref{sec:models}. 
In Sec.~\ref{sec:photo} we present the tDMRG results for the short-time dynamics of the local density and of the momentum distribution function following a photoexcitation, which we model by polarizing a single dimer in the center of the system. 
In Sec.~\ref{sec:relax} we discuss how to estimate the quasiparticle momentum distribution from the numerical tDMRG results and the computation of the quasiparticle relaxation rates from a linearized Boltzmann equation ansatz, followed by Sec.~\ref{sec:conclusions}, in which we summarize.
The considerations on the LBE are complemented by \cref{app:LBE}. 
In \cref{app:U0}, details on the effect of boundary conditions on the momentum distribution are discussed.
Finally, in \cref{appendix_entangler} details for the MPS computations at finite temperatures and for the estimate of the energy density of the excitation are presented.

\section{Polaronic order and effective model}
\label{sec:models}

In Ref.~\onlinecite{sotoudeh17_prb95_235150}, a tight-binding model for the strongly correlated electronic, spin, and lattice degrees of freedom of the three-dimensional manganite Pr$_{1-x}$Ca$_x$MnO$_3$ is developed. 
Models of this type have been described by Hotta.\cite{Hotta2004} 
The parameters of the model have been extracted from first-principles calculations.\cite{sotoudeh17_prb95_235150}
Because the simulation of the full quantum dynamics following a photoexcitation is out of reach for the three-dimensional material, we replace it by a fictitious one-dimensional manganite, which still exhibits the main properties of the three-dimensional system.
As Ref.~\onlinecite{sotoudeh17_prb95_235150} already details the derivation of the microscopic model in the three-dimensional case, we sketch the basic ideas here only briefly, and mention the differences leading to the 1D model treated here.

\subsection{Tight-binding model for manganites}
\label{sec:manganite-model}
\begin{figure}[t]
	\begin{center}
		\includegraphics[width=0.89\linewidth]{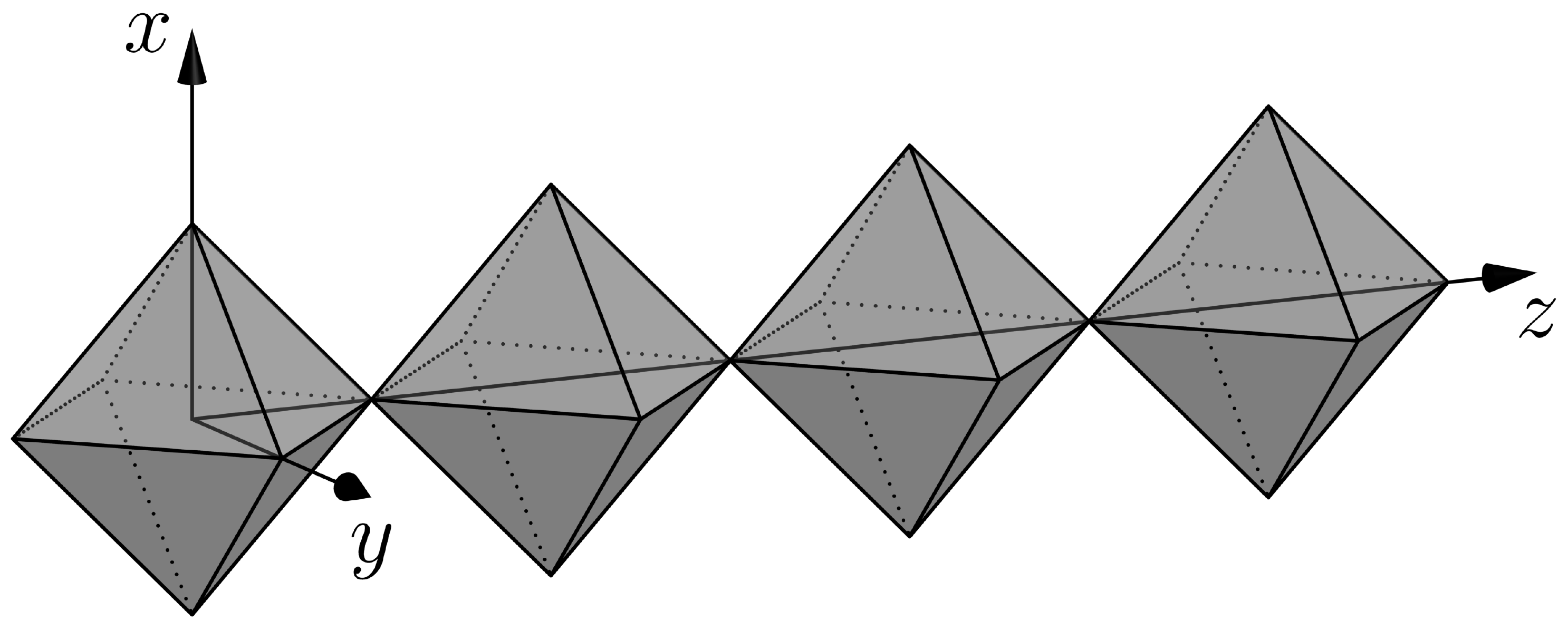}
	\end{center}
	\caption
	{
		\label{fig:octa_chain}
		One-dimensional chain of corner connected MnO$_6$ octahedra of the model manganite in the coordinate system chosen.
	}
\end{figure}
\begin{figure}[t]
	\begin{center}
	\includegraphics[width=0.99\linewidth]{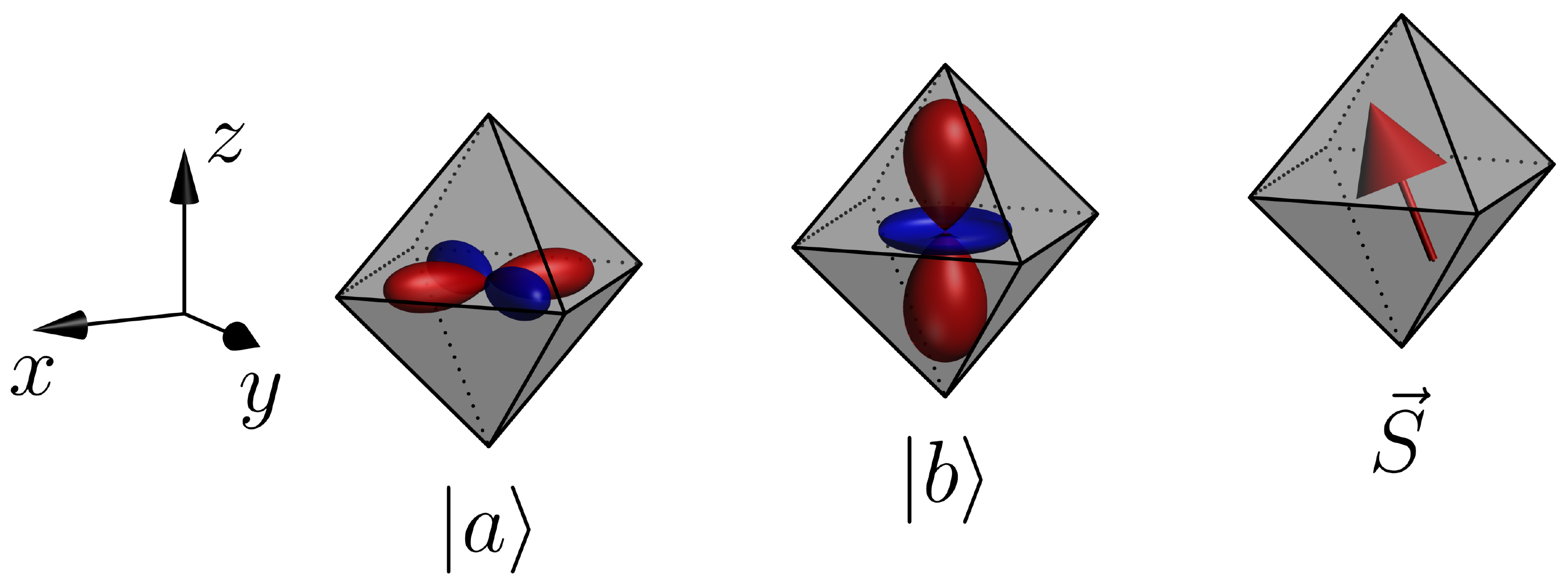}\\
	\includegraphics[width=0.99\linewidth]{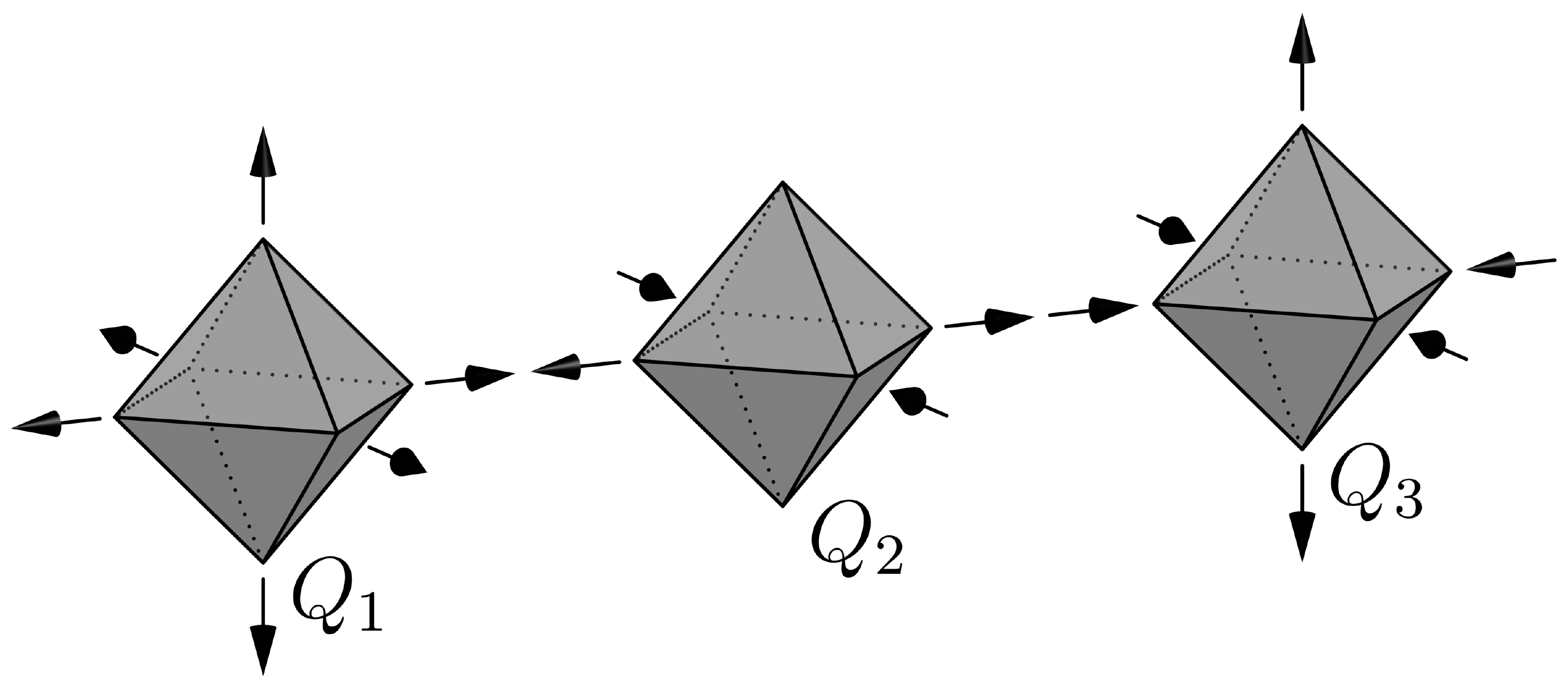}\\
	\end{center}
	\caption
	{
		\label{fig:octahedras}
		Degrees of freedom of the tight-binding model. 
		(top) Orbital degrees of freedom of the $e_g$ electrons,  which are treated explicitly, and classical spin degree of freedom related to the $t_{2g}$ electrons. 
		(bottom) Breathing mode $Q_1$ and Jahn-Teller-active phonon modes $Q_2$, $Q_3$.
	}
\end{figure}

The one-dimensional material consists of a chain of corner connected MnO$_6$ octahedra. 
The coordinate system has been chosen, as shown in \cref{fig:octa_chain}, with the $z$ axis along the chain, and the $x$ and $y$ axis along the orthogonal octahedral axes.

The dynamic electronic, spin and lattice degrees of freedom of the tight-binding model, sketched in \cref{fig:octahedras}, are the following:
\begin{itemize}
\item Electrons: The relevant electrons are those in the two $e_g$ orbitals of the Mn-ions. 
In the spirit of density-functional theory, the $e_g$ electrons are described by one-particle wave functions. 
The one-particle wave function with band index $n$ is expressed as 
\begin{equation}
	|\psi_n\rangle=\sum_{\sigma,\alpha,R}|\chi_{\sigma,\alpha,R}\rangle\psi_{\sigma,\alpha,R,n}
\end{equation} 
in terms of local spin-orbitals $|\chi_{\sigma,\alpha,R}\rangle$. 
The spin orbitals at Mn-site $R$ have spin $\sigma\in\{\uparrow,\downarrow\}$ and spatial orbital character $\alpha$, denoting the $d_{x^2-y^2}$ orbital for $\alpha=a$ and the $d_{3z^2-r^2}$ orbital for $\alpha=b$.

\item Spins: The three low-lying, spin-aligned electrons in the Mn-$t_{2g}$ states at site $R$ are described by a classical spin $\vec{S}_R$ of size $\frac{3}{2}\hbar$. 
Recent calculations indicate this to be an excellent approximation.\cite{hansgerd1}
\item Lattice: 
The relevant phonons are the two Jahn-Teller active distortions of the MnO$_6$ octahedra.
Their phonon amplitudes are denoted by $Q_{2,R}$ and $Q_{3,R}$.\cite{Dagotto2003}
The mode $Q_{3,R}$ describes the oblate and prolate distortion of the octahedron at site R along the $z$ direction. 
The mode $Q_{2,R}$ describes the simultaneous elongation along the $x$ and contraction along the $y$ direction, and vice versa. 
We refer here to the local Cartesian coordinates aligned along the octahedral axes. 
\end{itemize}
All other degrees of freedom are either absorbed into the dynamical variables of the model, or they are considered as a bath and not treated explicitly.

The total-energy functional of density-functional theory\cite{hohenberg64_pr136_864,kohn65_pr140_A1133} is replaced by the potential energy of the model, which takes the form
\begin{equation}
	\label{equ:pot-energy-fctal}
	E_{pot} = E_{e}+E_{S}+E_{ph}+ E_{e-ph} +E_{e-S}\;, 
\end{equation}
where $E_e$ is the energy of the electronic subsystem in the Mn-$e_g$ orbitals, $E_S$ describes the antiferromagnetic interaction of the $t_{2g}$ electrons on neighboring Mn sites, and $E_{ph}$ is the energy of the Jahn-Teller active phonons. 
The coupling of electrons with the spin of the $t_{2g}$ electrons and the lattice vibrations is described by  $E_{e-S}$ and $E_{e-ph}$, respectively.
We treat the different terms in the following way:
\begin{itemize}
\item Electronic energy $E_e$: 
The electronic energy contribution $E_e=E_{kin}+E_U$ consists of the kinetic energy $E_{kin}$ of the $e_g$ electrons and their interaction~$E_U$.
In the tight-binding model, the interaction $E_U$ is treated in a Hartree-Fock-like manner, analogous to hybrid density functionals\cite{becke93_jcp98_1372}, LDA+$U$\cite{anisimov91_prb44_943} and GW\cite{hedin65_pr139_796} calculations. 
All on-site matrix elements of the Coulomb interaction within a $e_g$ shell are considered. 
They can be parameterized by two Kanamori parameters\cite{Kanamori1963} $U$ and $J_{xc}$.
The hopping amplitude of the kinetic energy term is denoted as $\hopping$.
\item Spin energy $E_S$: 
The spin energy describes the coupling between the $t_{2g}$ states of neighboring Mn sites.
The coupling is antiferromagnetic and described by the parameter $J_{AF}$.
\item Phonon energy $E_{ph}$: 
The phonon energy describes a restoring force term which restores the perfect octahedron in the absence of other interactions. 
Note that the octahedral distortions of different sites are not independent but strongly coupled via oxygen atoms shared between two octahedra.
\item Electron-phonon coupling $E_{e-ph}$: 
The electron phonon coupling is responsible for the Jahn-Teller distortions of the octahedra in the presence of electrons.
\item Electron-spin coupling $E_{e-S}$: 
The electron spin coupling is responsible for the Hund's coupling $J_H$ between electrons in the Mn-$e_g$ orbitals, which are described explicitly, and the Mn-$t_{2g}$ electrons represented by classical spins.
This term is the origin of superexchange and double exchange, which are responsible for the complex magnetic properties of manganites.
\end{itemize}

The parameters of the model are extracted from \textit{ab initio} calculations of the three-dimensional manganites by using the projector-augmented wave method\cite{bloechl94_prb50_17953} in combination with the local hybrid density functional PBE0\(^{\text{r}}\) (for details, see Ref.~\onlinecite{sotoudeh17_prb95_235150}). 
Compared to the treatment in Ref.~\onlinecite{sotoudeh17_prb95_235150}, we include two changes: First, we ignore the breathing distortion $Q_1$ used in the original 3D model and, second, we increase the antiferromagnetic coupling for the one-dimensional model from 12~meV to 32.6~meV.
The latter was necessary to avoid a ferromagnetic ground state, while the 3D material exhibits a complex antiferromagnetic order.
The set of parameters obtained is reproduced in Table~\ref{tab:parameters}. 
\begin{table}[t]
	\caption
	{
		Model parameters for the one-dimensional model situation, based on the first-principle calculations on Pr$_{1-x}$Ca$_{x}$MnO$_3$ of Ref.~\onlinecite{sotoudeh17_prb95_235150}. 
		$J_{AF}$ describes the antiferromagnetic coupling between the $t_{2g}$ states of neighboring Mn sites; $J_H$ is the Hund's coupling; $U$ and $J_{xc}$ are the Kanamori parameters for electron-electron interaction between $e_g$ electrons; $g_{JT}$ and $k_{JT}$ parametrize the electron-phonon interaction; $\hopping$ is the hopping amplitude of the $e_g$ electrons.
		With two exceptions described in the text, they are identical to the values extracted in Ref.~\onlinecite{sotoudeh17_prb95_235150}. 
	}
	\begin{center}
		\begin{tabular}{lrllrl}
			\hline
			\hline
			$J_{AF}$  				& 32.6&meV \qquad\qquad\qquad & $g_{JT}^{\whitedagger}$	& 2.113&eV/\AA			\\
			$J_H$    				& 0.653&eV & $k_{JT}$  			& 5.173&eV/\AA\(^2\)	\\
			$U$      				& 2.514&eV & $t_{hop}$ 			& 0.585&eV		\\
			$J_{xc}$  				& 0.692&eV &                            &&\\
			\hline
			\hline
		\end{tabular}
	\end{center}
	\label{tab:parameters}
\end{table}

\subsection{Polaron and magnetic order in 1D manganites}
\label{sec:MD-results}

In this section, we describe the ground-state configurations of the one-dimensional manganite chain as obtained from the tight-binding model Eq.~\eqref{equ:pot-energy-fctal}. 
Beyond the purpose of  providing the ground states of the one-dimensional model, the motivation for the present study has been to explore how the complex phase diagram and the polaron arrangement of manganites in general can be described and analyzed. 
The polaron model derived in the following is a promising approach towards this goal.

\begin{figure}[b]
	\begin{center}
		\includegraphics[width=0.49\linewidth]{cmo_s}
		\includegraphics[width=0.49\linewidth]{pmo_s}\\
		\vspace{3mm}
		\includegraphics[width=0.49\linewidth]{electron_trimer_s}
		\includegraphics[width=0.49\linewidth]{hole_trimer_s}\\
		\vspace{3mm}
		\includegraphics[width=0.49\linewidth]{zener_s}
		\includegraphics[width=0.49\linewidth]{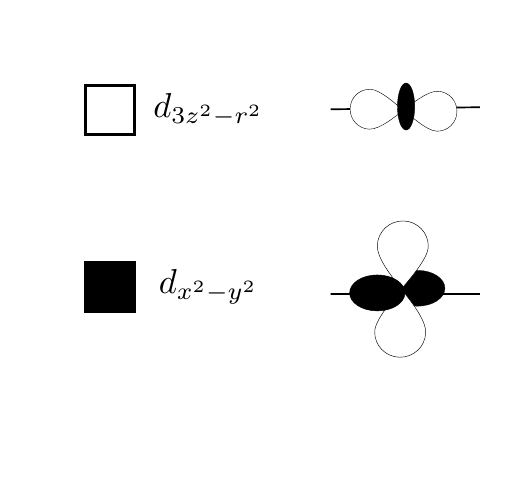}
	\end{center}
	\caption
	{
		\label{fig:polarons}
		Projected density of states of the tight-binding model for the polarons in the 1D manganite chain as a function of energy in eV. 
		Top left: Two adjacent unoccupied sites $V$; top-right: two adjacent Jahn-Teller polarons $P^{JT}$; middle left: electron polaron $P^e$; middle right: hole polaron $P^h$; bottom Zener polaron $P^Z$. 
		The horizontal line indicates the Fermi level. 
		Empty and filled y lines indicate $d_{3z^2-r^2}$ orbitals pointing along the chain and $d_{x^2-y^2}$ orbitals orthogonal to the chain, respectively.
		The density of states is broadened by 0.05~eV.
	}
\end{figure}

The ground states have been determined in a two step approach: 
First, stable and metastable configurations are obtained using Car-Parrinello-like dynamics \cite{car85_prl55_2471} with friction.
\footnote{We have done calculations for different unit cells. 
We choose a twelve-site model because it produces only a few frustrated structures ($N_s$ must be divisible by $2$, $3$, and $4$.) 
The size of the unit cell has been chosen small to avoid being trapped in metastable states.} 
Second, the emerging patterns and their building blocks are identified.
Moving towards a higher-level description, the total energy is then expressed in terms of the energies of these building blocks.
The magnetic order and the resulting polaron composition are presented in Table~\ref{tab:orders}.
The electronic structure, i.e., the density of states, of the various polaron types is provided in Fig.~\ref{fig:polarons}. 

\begin{table*}[th!]
	\caption
	{
		\label{tab:orders}
		Magnetic orders, polaron composition, deviation $E_{PM}-E_{TB}$ of the energy $E_{PM}$ from the polaron model, Eq.~\eqref{eq:polaronetot}, and energy $E_{TB}$ from the tight-binding model Eq.~\eqref{equ:pot-energy-fctal} for different numbers of electrons in a 12-site unit cell of the 1D manganite. 
	}
	\begin{center}
		\begin{tabular}{|c|c|c|c|c|}
			\hline
			\hline
			$N_e$ &  magnetic order & composition  & $(E_{TB}-E_{PM})$[meV] & $E_{TB}$[eV]\\
			\hline
			0& $\uparrow \downarrow \uparrow \downarrow \uparrow \downarrow \uparrow \downarrow \uparrow \downarrow \uparrow \downarrow$ & $V_{12}$ &0 & -0.3912\\
			1& $ \downarrow\uparrow\uparrow\uparrow\downarrow \uparrow \downarrow \uparrow \downarrow \uparrow \downarrow \uparrow $ & $P^eV_9$ & 0  &  -1.98129 \\
			2& $\downarrow \uparrow \uparrow \uparrow \downarrow \uparrow \uparrow \uparrow\downarrow \uparrow \downarrow \uparrow$ & $(P^eV)_2V_4$ & 0.5 & -3.57088 \\
			3& $\downarrow \uparrow \uparrow \uparrow \downarrow \uparrow \uparrow \uparrow\downarrow \uparrow \uparrow \uparrow$ & $(P^eV)_3$ & 10.7 & -5.15075  \\
			4& $\uparrow \uparrow \uparrow \downarrow \downarrow \downarrow\uparrow \uparrow \uparrow \downarrow \downarrow \downarrow $ &$P^e_4$ & 59.3 & -6.69231\\
			5& $^{*1}$  noncollinear & $P^Z_3P^e_2$  & -16.2 & -8.08389\\
			6& $\uparrow \uparrow \downarrow \downarrow\uparrow \uparrow  \downarrow \downarrow \uparrow \uparrow \downarrow \downarrow $ &$P^Z_6$ & 0 & -9.38383\\
			7 & $\uparrow \uparrow \downarrow \downarrow \uparrow \downarrow \downarrow  \uparrow \uparrow\downarrow \downarrow \downarrow $ &$P^Z_4P^hP^{JT}$ & -46.7 & -10.46353 \\
			8 & $^{*2} \uparrow \uparrow \uparrow \downarrow \downarrow \uparrow \downarrow \uparrow \uparrow  \downarrow \uparrow \uparrow $ &$P^Z_3P^hP^{JT}_3$& -12.0 & -11.51915 \\
			9& $^{*3} \uparrow \uparrow \downarrow \uparrow \uparrow\downarrow \uparrow \downarrow  \uparrow \uparrow \uparrow  \downarrow $ & $P^z_2P^hP^{JT}_5$ & -68.9 & -12.51556 \\
			10& $\uparrow \downarrow \uparrow \downarrow \downarrow \uparrow \downarrow \uparrow \uparrow \downarrow \uparrow \downarrow $ &$P^z_2P^{JT}_8$ & -74.4 & -13.51869 \\
			11& $\downarrow \uparrow \uparrow \uparrow \downarrow \uparrow \downarrow \uparrow \downarrow \uparrow \downarrow \uparrow$ &$P^hP^{JT}_9$ & 0 & -14.47726 \\
			12& $\uparrow \downarrow \uparrow \downarrow \uparrow \downarrow \uparrow \downarrow \uparrow \downarrow \uparrow \downarrow$ &$P^{JT}_{12}$ & 0 & -15.47453\\
			\hline
		\end{tabular}
	\end{center}
	$*1$- All the angles between classical spin vectors lie in the range of $\sim (162-175)^o$. \\
	$*2$- The average angle within the trimer is $\sim < 51^o>$ and  other angles are in the range $\sim (157-166)^o$.\\
	$*3$ - The average angle within the trimer is $\sim<39.5^o>$ and  other angles are in the range $\sim (162-175)^o$.\\
\end{table*}
				
We obtained the following dominant patterns, which we describe as polarons:
\begin{enumerate}
\item There are sites $V$ without $e_g$ electrons, which we denote as vacant sites.
The spins on these sites interact only weakly with each other via the antiferromagnetic Heisenberg exchange coupling $J_{AF}$.
In this language, the electron-poor manganite -- analogous to CaMnO$_3$ -- consists of tightly packed $V$ sites.
\item The electron polaron $P^e$ is a trimer of ferromagnetically aligned Mn sites occupied by a single $e_g$ electron. 
$P^e$ is analogous to an electron polaron in CaMnO$_3$.
$P^e$ has three electron states in the majority-spin direction: 
The lowest state is occupied and fully bonding. 
The second state is unoccupied and nonbonding. 
This state is distributed over the two outer sites of the trimer.  
The third state of the electron polaron is fully antibonding.
\item The Zener polaron\cite{zener51_pr82_403,zhou00_prb62_3834} $P^Z$ is a dimer of ferromagnetically aligned Mn sites, which share a single $e_g$ electron. 
The half-doped 1D material -- analogous to Pr$_{1/2}$Ca$_{1/2}$MnO$_3$ (PCMO) -- can be described as a crystal of antiferromagnetically coupled Zener polarons. 
The Zener polaron has two states in the majority spin direction: a filled bonding and an empty antibonding state. 
\item The hole polaron $P^h$ is a trimer of ferromagnetically aligned Mn sites occupied by two $e_g$ electrons. 
$P^h$ is analogous to a hole polaron in PrMnO$_3$.
It has the same three states as the electron polaron, however the second, nonbonding state is occupied as well.
\item The Jahn-Teller polaron $P^{JT}$ is an $e_g$ electron that occupies a single site. 
A crystal of Jahn-Teller polarons is analogous to PrMnO$_3$.
\end{enumerate}

In order to extract the energies for these structural units, we start out by setting the reference $\mu_0$ for the electron chemical potential to the coexistence value of the electron-poor $(N_e=0)$ and the electron-rich $(N_e=N_s)$ systems that are analogous to CaMnO$_3$ and PrMnO$_3$, respectively. 
That is, instead of the energy $E$, we consider the energy $E-\mu_0 N_e$ of the manganite together with a conveniently chosen electron reservoir.

\begin{figure}[h!]
	\begin{center}
		\includegraphics[width=\linewidth]{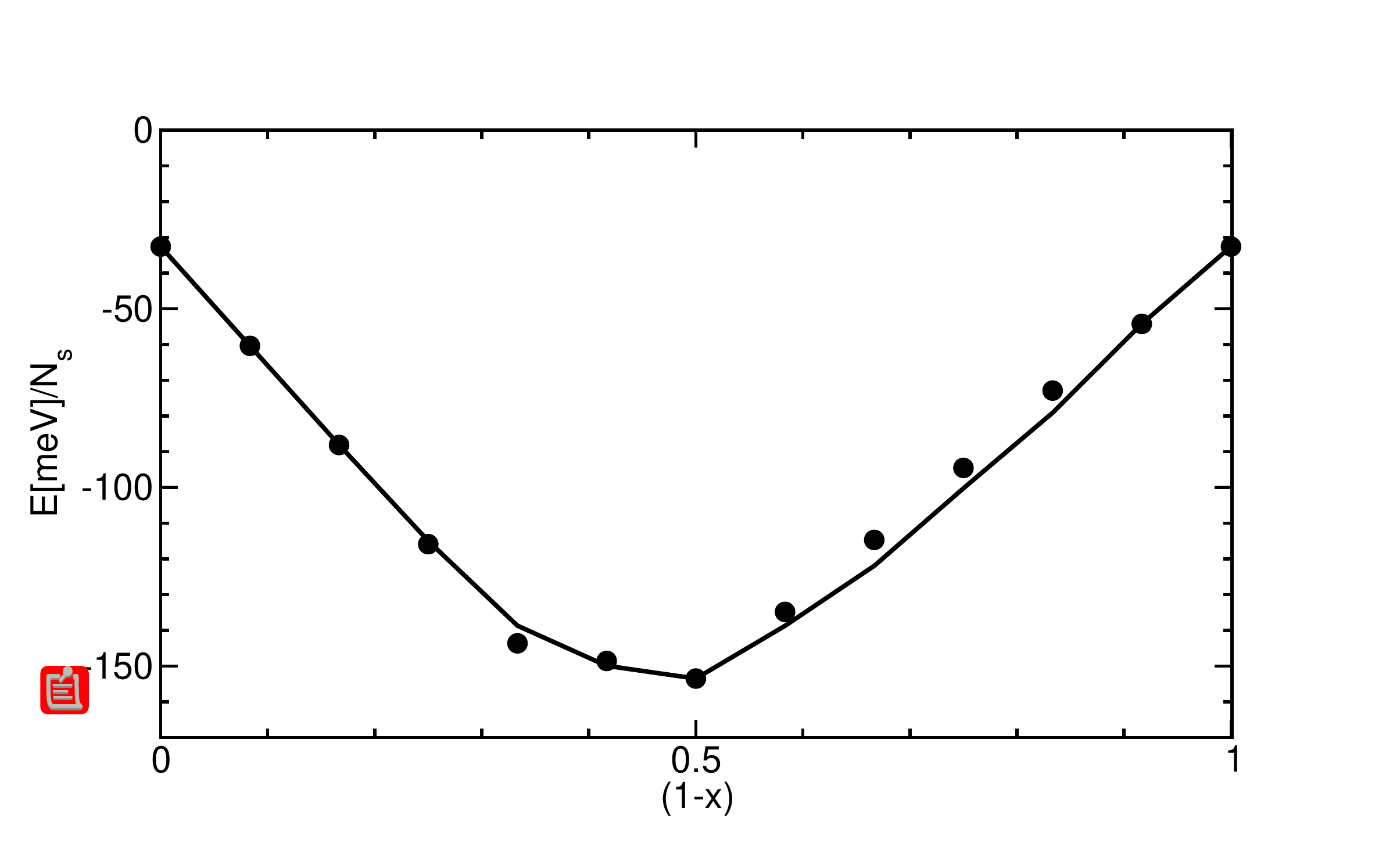}
	\end{center}
	\caption
	{
		\label{fig:polaronsumenergy}
		Energy per Mn site of the model calculation (line) as a function of the electron occupation $1-x=N_e/N_s$ compared to the sum of polaron energies given in Eq.~\eqref{eq:polaronetot} (symbols). 
		The energy $(1-x)\mu_0$ of the particle reservoir has been included. 
	}
\end{figure}

With $N_s$, we denote the number of sites in the unit cell and with $N_e$ the number of electrons per unit cell.
Then, we identify the polaron composition from the magnetic order. 
The formation energies of the polarons are determined in such a way that the energy 
\begin{equation}
	\label{eq:polaronetot}
	\begin{split}
		E_{PM}[n_V,n_e,n_Z,n_h,n_{JT}]= \\
		\sum_{j\in\{e,Z,h,JT\}} n_j (E_f^{(j)}+\mu_0 N^{(j)}_e-J_{AF})
	\end{split}
\end{equation}
of the polaron model matches the total energies obtained from the tight-binding calculation given in Table~\ref{tab:orders}.
With $n_j$, we denote the number of polarons $P^{(j)}$, where $j\in\{V,e,Z,h,JT\}$ denotes the polaron type.
With $E_f^{(j)}$, we denote the polaron formation energy and with $N^{(j)}_e$ the number of $e_g$ electrons in the respective polaron. 
The values are presented in Table~\ref{tab:polaronenergies}. 
Specifically, the polaron-formation energies have been extracted so that the energies of the model calculations are reproduced by Eq.~\eqref{eq:polaronetot} for $N_e=0,1,6,11,$ and $12$. 
Figure~\ref{fig:polaronsumenergy} compares the energetics of the configurations in Table~\ref{tab:orders} with that of the polaron model, Eq.~\eqref{eq:polaronetot}, using the values from Table~\ref{tab:polaronenergies}. 
It is evident that the simple polaron model captures the major features of the ground state energetics.

\begin{table}[!htb]
	\caption
	{
		\label{tab:polaronenergies}
		Formation energies of polarons, number $N_s^{(j)}$ of sites occupied by the polaron $P^{j}$, and number $N_e^{(j)}$ of $e_g$ electrons on it.
	}
	\begin{tabular}{|l|ccccc|}
		\hline
		\hline
		polaron & V & $P^e$ &   $P^Z$       &  $P^h$   &   $P^{JT}$   \\
		\hline
		$E_f^{(j)}$[meV]   & 0 & -398.3 & -274.4 & -324.9  &  0 \\
		$N_s^{(j)}$     & 1 & 3        & 2        & 3         & 1\\
		$N_e^{(j)}$     & 0 & 1        & 1        & 2         & 1\\
		\hline
		\hline
	\end{tabular}
\end{table}
Interestingly, we can attribute a definite size to the electron polaron:
The mechanism limiting the size of the electron polaron $P^e$ is the competition of the kinetic energy with the antiferromagnetic coupling.
Increasing the size of the electron polaron on one side lowers the kinetic energy of the electron, because it can spread over a larger region.
On the other side, there is a penalty for aligning more sites ferromagnetically. 
The maximum size of the electron polaron is reached when the delocalization energy gained by extending the electron polaron by one site is exceeded by the antiferromagnetic coupling. 
In the limit of large Hund's coupling $J_H$, the size of the polaron is thus determined by the ratio $t_{hop}/J_{AF}$ of hopping parameter and antiferromagnetic coupling. 
With our set of parameters, this maximum size is three sites.

In the dilute limit, i.e., for $N_e\le3$, we find that the polarons are separated by at least one vacant site $V$, as if there were a nearest-neighbor repulsion between adjacent polarons. 
We attribute this repulsion to the Coulomb interaction between the wave function tails, which extend into a neighboring polaron, with the electrons already belonging to this neighboring polaron.

Because of the kinetic energy cost, the smaller polarons, such as $P^Z$, are energetically less favorable than larger polarons, such as $P^e$. 
Thus, they become relevant only when the electron density is such that the larger polarons, namely the electron polaron $P^e$, are densely packed. 
For our system, this occurs at $N_e/N_s=1/3$.

Beyond this value, electron polarons $P^e$ and Zener polarons $P^Z$ coexist until Zener polarons are densely packed. 
This is the case for half doping, i.e., $N_e/N_s=1/2$. This is the doping used for the study of the optical excitation, which will be discussed in the following section.

In the electron-rich phase with $N_e/N_s=1$, the system forms a solid of antiferromagnetically coupled Jahn-Teller polarons $P^{JT}$.

When doping the electron-rich phase with holes, the preferred way is via a Zener polaron $P^Z$.
In analogy with the electron-poor manganite, one would have expected that the extended defect $P^h$ was favorable compared to the smaller Zener polaron $P^Z$.
The reason for the preference of the Zener polaron is that the formation of hole polarons $P^{h}$ from Zener and Jahn-Teller polarons requires substantial energy
\begin{eqnarray}
	\label{eq:formationholepolarons}
	P^{JT}+P^{Z}\rightarrow P^h-50.5~\text{meV} \; .
\end{eqnarray}

Nevertheless, we encounter hole polarons in our calculations. 
They are formed in response to spin frustration.
The insertion of a hole into the electron rich material by forming a Zener polaron would, at the same time, introduce a domain wall into the antiferromagnetic order. 
An isolated domain wall can either annihilate with another domain wall or combine with a Zener polaron to form a hole polaron.
With the hole polaron, we identified a structural unit that does not contribute to the ground state at zero Kelvin, but that plays an important role for the interconversion of polarons.

The nature of an isolated domain wall can be rationalized from the point of view of a Zener polaron. 
An abrupt domain wall in the electron-rich material is equivalent to a Zener polaron with one additional electron. 
The additional electron enters into an antibonding state, which is energetically highly unfavorable. 
By forming a hole polaron, this electron is transferred into the nonbonding state of the hole polaron, which is energetically favorable.

The occurrence of noncollinear spin arrangements indicates that the phase boundary can also delocalize and form a spin spiral. 
The energy scale of forming these polarons is of the order of 0.3~eV, while that of the interaction between polarons is much smaller, i.e., of the order of 10~meV.

There is an analogy of the description of the order in manganites in terms of various types of weakly interacting polarons with molecules in chemistry. A polaron is the analogon of a molecule. 
Electrons that delocalize over several Mn-sites are analogous to a chemical bond. 
Similar to molecules, which can arrange into molecular crystals, the polarons arrange in various patterns, which give rise to the complex phase diagram of manganites.
The conversion of polarons into each other is then  analogous to a chemical reaction.
An example for such a reaction between polarons is the formation of a hole polaron from a Jahn-Teller and a Zener polaron in Eq.~\eqref{eq:formationholepolarons} described above. 
%

\begin{figure*}
	\begin{center}
		\includegraphics[trim=3 0 0 0,clip,width=0.95\textwidth]{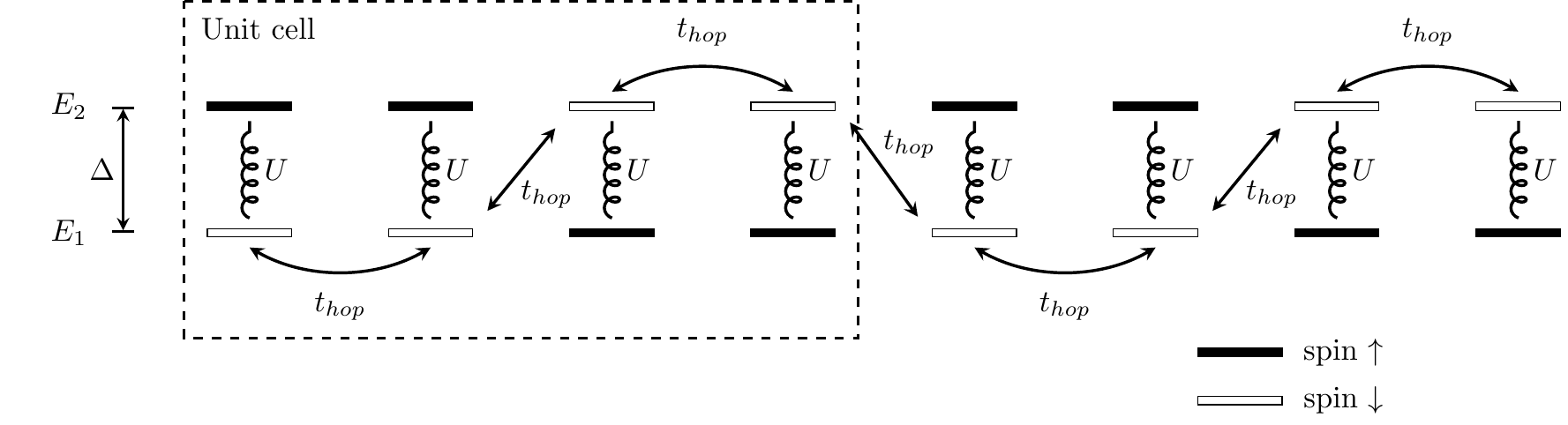}
		\caption
		{
			\label{fig: sketch eff model}
			Sketch of the effective Hubbard-type many-body model derived in this section. 
			The unit cell consists of four sites, with each site hosting a spin-up and a spin-down state. 
			The two states are energetically separated by the Hund's splitting $\Delta= 2 J_H$. 
			Electrons can hop from one site to another with the hopping amplitude $\hopping$, provided the two states have the same spin. 
			For the sake of clarity this is only shown for the spin-$\downarrow$ direction.
			If two electrons are located on the same site, they feel the screened Coulomb repulsion $U$ (depicted by the curly line). 
			Otherwise, this repulsion is screened by the positively charged atoms.	
		}
	\end{center}
\end{figure*}

\subsection{Hamiltonian for a frozen lattice of Zener polarons}
\label{sec:1D-model-valence-el}
In the previous section, we explored the ordered phases of the 1D model manganite at low temperatures described by Eq.~\eqref{equ:pot-energy-fctal}. 

In order to study the light-absorption process and the electronic relaxation, we focus on the electronic degrees of freedom. 
In the following, we therefore freeze the spin and lattice degrees of freedom in the ground state. 
The only dynamical entities in this model are the $e_g$ electrons. 
Furthermore, the Hilbert space for the $e_g$ electrons has been limited to two $d_{3z^2-r^2}$ spin orbitals per Mn site, i.e., $|\chi_{\sigma,b,R}\rangle$, which makes the model similar to a single-band Hubbard model with spatially varying magnetic fields.

Furthermore, we will focus on the half-doped system, because it allows us to study the role of the magnetic microstructures formed by antiferromagnetically coupled Zener polarons on the relaxation dynamics of a photoexcitation. 
This order corresponds to $N_e=6$ in Table~\ref{tab:orders}.

As shown in \cref{fig:polarons}, such a Zener polaron consists of two neighboring Mn sites, which share a single $e_g$ electron that is uniformly delocalized over both sites. 
The Mn ions inside a Zener polaron are ferromagnetically aligned, and, without loss of generality, we choose the spins to point along the $z$ axis, that is $S_x=S_y=0$.
This leads to the spin configuration on the four Mn-sites of the unit cell, 
\begin{eqnarray}
	\Bigl(S_{z,1},S_{z,2},S_{z,3},S_{z,4}\Bigr)=\frac{3\hbar}{2}\Bigl(-1,-1,+1,+1\Bigr) \, .
	\label{eq:Szr}
\end{eqnarray}
The spin distribution is periodic, so that  $S_{z,R+4}=S_{z,R}$.
This means that the electrons experience the spin and lattice degrees of freedom as a staggered magnetic field. 

Because of the restriction to a collinear spin distribution of the classical spins $\vec{S}_R$, the two spin directions of the electrons decouple, except through the electron-electron interaction.

Since the Jahn-Teller distortions do not modulate the potential for the remaining orbitals of the state under investigation, we also omit the electron-phonon coupling.
As a result, the total energy can be expressed in the form of a one-band Hubbard model with a staggered magnetic field and total energy
\begin{eqnarray}
	E=E_{kin}+E_{U}+E_{e-S} \; .
\end{eqnarray}

Formulated in second quantization, we thus obtain the simplified many-electron Hamiltonian for a half-doped 1D manganite,  
\begin{eqnarray}
	\label{equ: eff model Hamiltonian}
	\hat{H}=\sum_{R}&\biggl\lbrace&	
	-t_{hop}\sum_\sigma \Bigl( \hat{c}^\dagger_{\sigma,R+1}\hat{c}^{\phantom{\dagger}}_{\sigma,R}+ \hat{c}^\dagger_{\sigma,R}\hat{c}^{\phantom{\dagger}}_{\sigma,R+1}\Bigr) \nonumber\\
	&+&U \hat{n}_{\uparrow,R}\hat{n}_{\downarrow, R} 
	+ \frac{\Delta}{3\hbar}S_{z,R} \left(\n_{\uparrow,R}-\n_{\downarrow,R} \right)
	\biggr\rbrace \; ,
\end{eqnarray}
with $\hat c^{(\dagger)}_{\sigma, R}$ the annihilation (creation) operator for an electron of spin $\sigma$ at position $R$, and the local spin occupation $\hat{n}_{\sigma,R}:=\hat{c}^\dagger_{\sigma,R}\hat{c}^{\phantom{\dagger}}_{\sigma,R}$.

Using the values of Table~\ref{tab:parameters}, we obtain
\begin{equation}
	U \approx 4.3 t_{hop}
	\label{eq:U}
\end{equation}
for the Hubbard interaction and 
\begin{equation}
	\Delta:=2{J_H}\approx 2.3 t_{hop}\;
	\label{eq:Delta}
\end{equation}
for the Hund's splitting.

The resulting Hubbard-type model thus has a unit cell of four sites and is sketched in Fig.~\ref{fig: sketch eff model}. 
In relation to PCMO, it will be interesting to study the photoexcitation for the set of parameters \eqref{eq:U} -- \eqref{eq:Delta}. 
However, model Eq.~\eqref{equ: eff model Hamiltonian} allows us to go beyond and tune the values of $U/\hopping$ and $\Delta/\hopping$ independently from each other.
Consequently, this model realizes a minimal model for a manganite system to study the interplay between the Hund's coupling and the electron-electron interaction after a photoexcitation in such systems. 
In the following, we will hence study the time evolution after a photoexcitation for the parameter values \eqref{eq:U} -- \eqref{eq:Delta} using MPS and LBE techniques, and also the results when changing the values of $\Delta/t_{hop}$ and $U/t_{hop}$. 

\subsubsection{Band structure of non-interacting electrons in the lattice of frozen Zener polarons}
\label{sec:bandstructure}

Before discussing the photoexcitations, let us first explore the basic features of model Eq.~\eqref{equ: eff model Hamiltonian} without Coulomb interaction, i.e., the case $U=0$.

The band structure of the noninteracting system will elucidate the role of the Hund's splitting $\Delta$, which acts as a staggered magnetic field on the electronic structure. 
We obtain 
\begin{eqnarray}
	\epsilon_\nu(k) &=& s_{1,\nu} \hopping\sqrt{2+\tilde{\Delta}^2 +s_{2,\nu} 2 \sqrt{\cos^2(2ka)+\tilde{\Delta}^2}} \; , \nonumber\\
	\label{equ:1p-band-structure}
\end{eqnarray}
where $\tilde{\Delta} = \frac{\DelX}{2\hopping}$, $k$ is the momentum in the reduced Brillouin zone, $\nu$ labels the bands in this reduced Brillouin zone, and $(s_{1,\nu},s_{2,\nu})=(-1,+1), (-1,-1), (+1,-1)$ and $(+1,+1)$ for $\nu=1,2,3,4$. 
The spacing between the Mn ions is denoted by $a$.
For the details of the derivation, see \cref{app:LBE}.

\begin{figure}[t]
	\centering
	\includegraphics[width=1\linewidth]{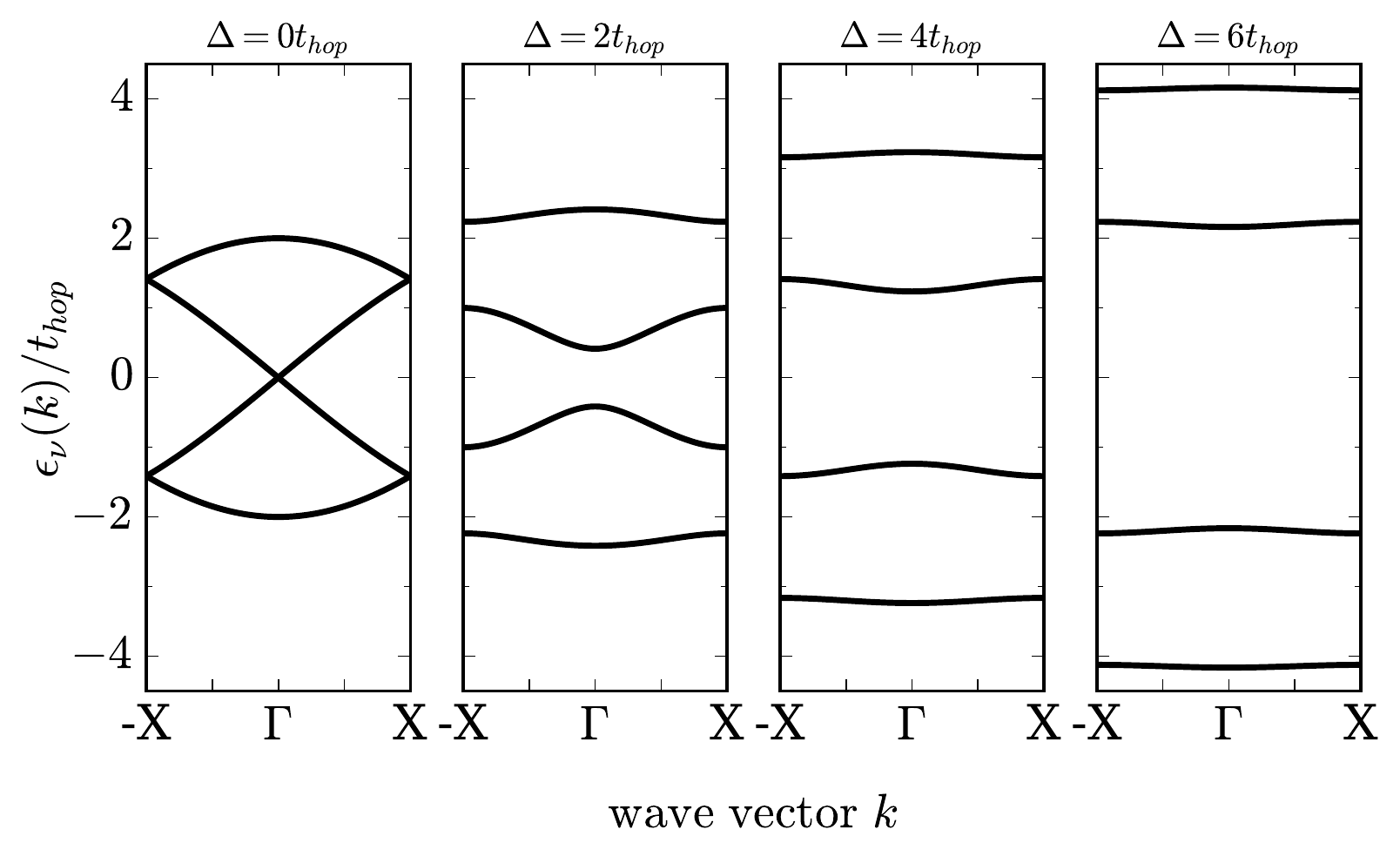}
	\caption
	{
		\label{fig:bandstructure}
		One-particle band structure of PCMO for different values of the Hund's splitting $\Delta$, which is measured in units of $\hopping$.
		$\Gamma$ denotes the origin of the $k$ points and $\mathrm{X} =\pi/4a$ the zone boundary with the $\mathrm{Mn}$-$\mathrm{Mn}$ spacing $a$.
		One can see that the distance between the center of the upper two bands and the center of the lower two bands is close to $\Delta$ for large values of $\Delta$.
		Furthermore, in the same limit, the distance of the upper two bands (as well as the one of the lower two bands) is approximately $2\hopping$.
	}
\end{figure}

In \cref{fig:bandstructure}, this band structure is shown for different values of $\Delta/t_{hop}$. 
Without Hund's splitting, the system is equivalent to a single-band Hubbard chain, which has the band structure 
\begin{eqnarray}
	\epsilon(k)=-2t_{hop}\cos(ka) \, .
\end{eqnarray}
In the setting of the four-site unit cell, this band structure is folded back twice into the smaller reciprocal unit cell as shown in \cref{fig:bandstructure}. 
The lowest of the four bands is occupied.

In the limit of infinite Hund's splitting $\Delta$, the band structure develops into four nearly dispersionless bands. 
The nature of the states in this limit can be identified with those of a Hubbard dimer, respectively a hydrogen molecule. 
The isolated Hubbard dimer has a bonding and an antibonding state for each spin direction. 
The energetic separation of the two bands is given by the hopping parameter as $2\hopping$.  
The electrons of one spin direction experience a downward Zeeman-like shift by $J_H=\frac{1}{2}\Delta$ on one Zener polaron and a similar upward shift on the other Zener polaron. 
The resulting states are at $-\frac{1}{2}\Delta\pm \hopping$ and $\frac{1}{2}\Delta\pm \hopping$.

The bandwidth of the four bands is given by the ability of electrons to tunnel between two second-nearest neighbor Zener polarons, which have the same spin orientation.
The tunneling probability in turn acts as an effective hopping for the molecular states.

As seen in Fig.~\ref{fig:bandstructure}, intermediate Hund's splitting leads to a coexistence of gaps, flat bands, and bands with large dispersion.
Thus, the behavior of the band structure will be non-trivial and probably most interesting for intermediate Hund's splitting. 
The parameters in Table~\ref{tab:parameters} show that PCMO lies in this regime.

\section{Photoexcitation dynamics}
\label{sec:photo}

\subsection{Treatment of light-matter interaction}
\label{sec:light-exc}

A simple way to model the photoexcitation is to assume it to create particle-hole like excitations.\cite{Silva2010, PhysRevB.85.205127, Wall2011, PhysRevLett.111.016401, Dagotto2008, PhysRevLett.111.016401} 
Here, we start from a ground state described in terms of Zener polarons, in which the electron density is equally distributed.  
We then model the photoexcitation as inducing an electric dipole on a single polaron.
The conceptually simplest operator then is 
\begin{equation}
	\label{eq:single_exciton}
	\hat Y_R = \sum_{\sigma} \hat c^{\dagger}_{\sigma, R+1} \hat c^{\phantom{\dagger}}_{\sigma, R}\;,
\end{equation}
with $R$ and $R+1$ being lattice sites both located on the same polaron.  

In this paper, we will treat a single excitation on lattices with typically 40 sites. 
As discussed in \cref{finite_temp_app}, assuming a light pulse of duration of $1$~fs, this corresponds to an intensity of $\sim 10^8$~W/mm.

\subsection{Details of the tDMRG calculations}
\label{sec:detailstDMRG}

We use the two-site time evolution matrix-product operator (MPO) introduced in Ref.~\onlinecite{Zaletel2015} with a time step of \(\Delta t=0.05\).
In this approach, the operator exponential of the propagator
\begin{equation}
	\label{eq:time-ev-op}
	\hat U(t) = e^{-\frac{i}{\hbar} \hat H \Delta t}
\end{equation}
is given by an MPO. 
In Ref.~\onlinecite{Zaletel2015} two representations are introduced; we chose the one denoted as \(\hat W^{II}\), which is considered to be more accurate. 
The Hamiltonian is given in terms of finite state machines and is subsequently transformed into the MPO form. \cite{PhysRevA.78.012356,SciPostPhys.3.5.035}

Two main error sources need to be considered:
First, the error due to the truncation of the MPS matrices to dimension $\chi_\text{MPS}$.
In all simulations, the entanglement induced by the perturbation, as quantified by the von Neumann entropy,\cite{Amico:2008en} is rather small. 
Therefore, a matrix dimension of \(\chi_\text{MPS} = 512\) for systems with up to $L=40$ lattice sites was sufficient to obtain a discarded weight $\varepsilon \sim 10^{-8}$ ($\sim 10^{-4}$) at the end of the time evolutions for \(\Delta/\hopping=8\) (\(\Delta=0\)). 
The second source of error is due to the approximation of the operator exponential and is explained in detail in Ref.~\onlinecite{Zaletel2015}.
As it is much smaller than the error due to the truncation, this error is negligible.

The MPS code used is implemented using the SciPAL\citep{scipal} library, which is a framework based on C++ expression templates, and provides the possibility to use CPUs as well as GPUs by calling efficient implementations of BLAS and cuBLAS\footnote{{c}uBLAS is an implementation of the BLAS library especially for the usage on NVIDIA CUDA devices.} functions. 

\subsection{Short-time dynamics after the photoexcitation}
\label{subsec:DRMG-results}

\begin{figure*}[t]
	\includegraphics[width = 0.48\textwidth]{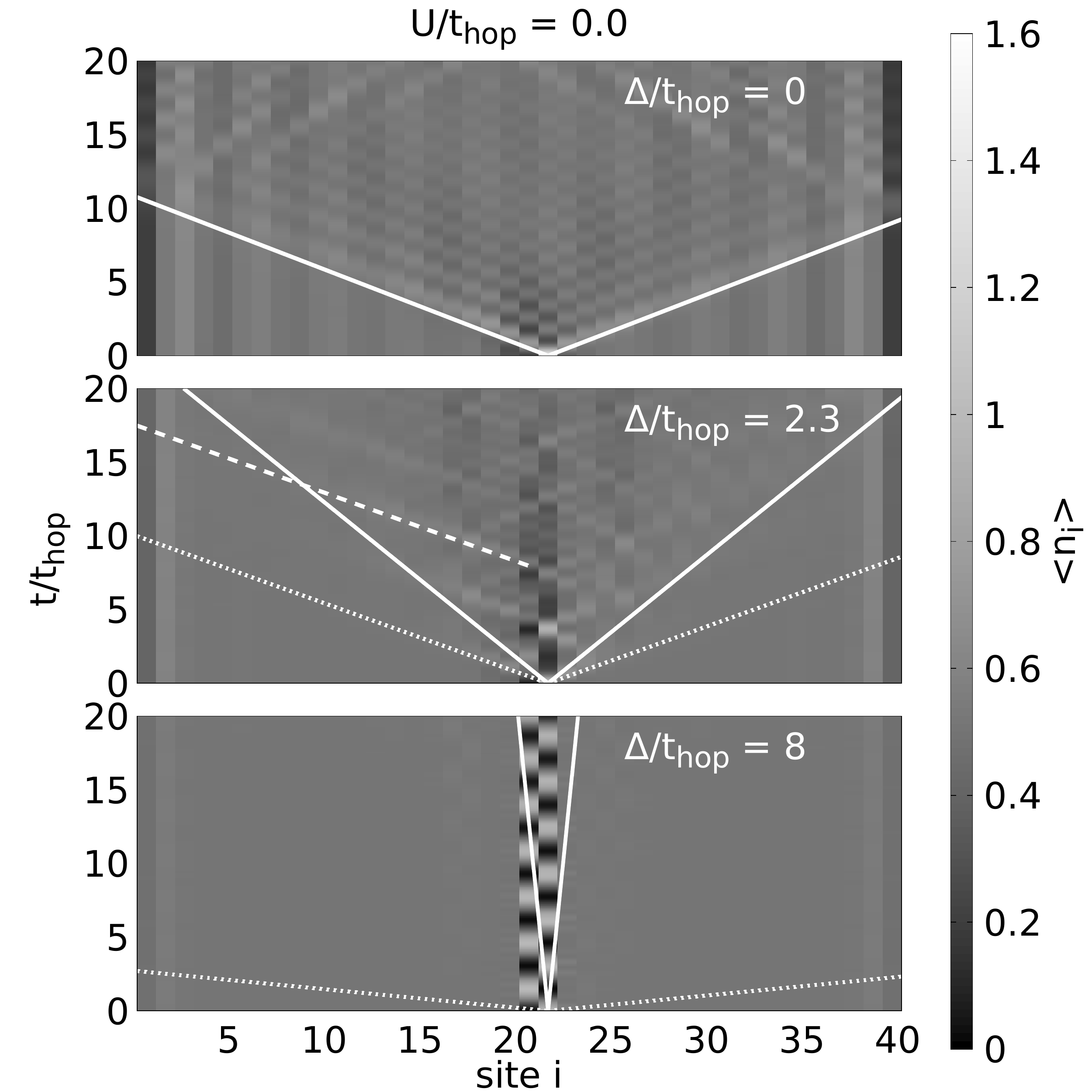}
	\includegraphics[width = 0.48\textwidth]{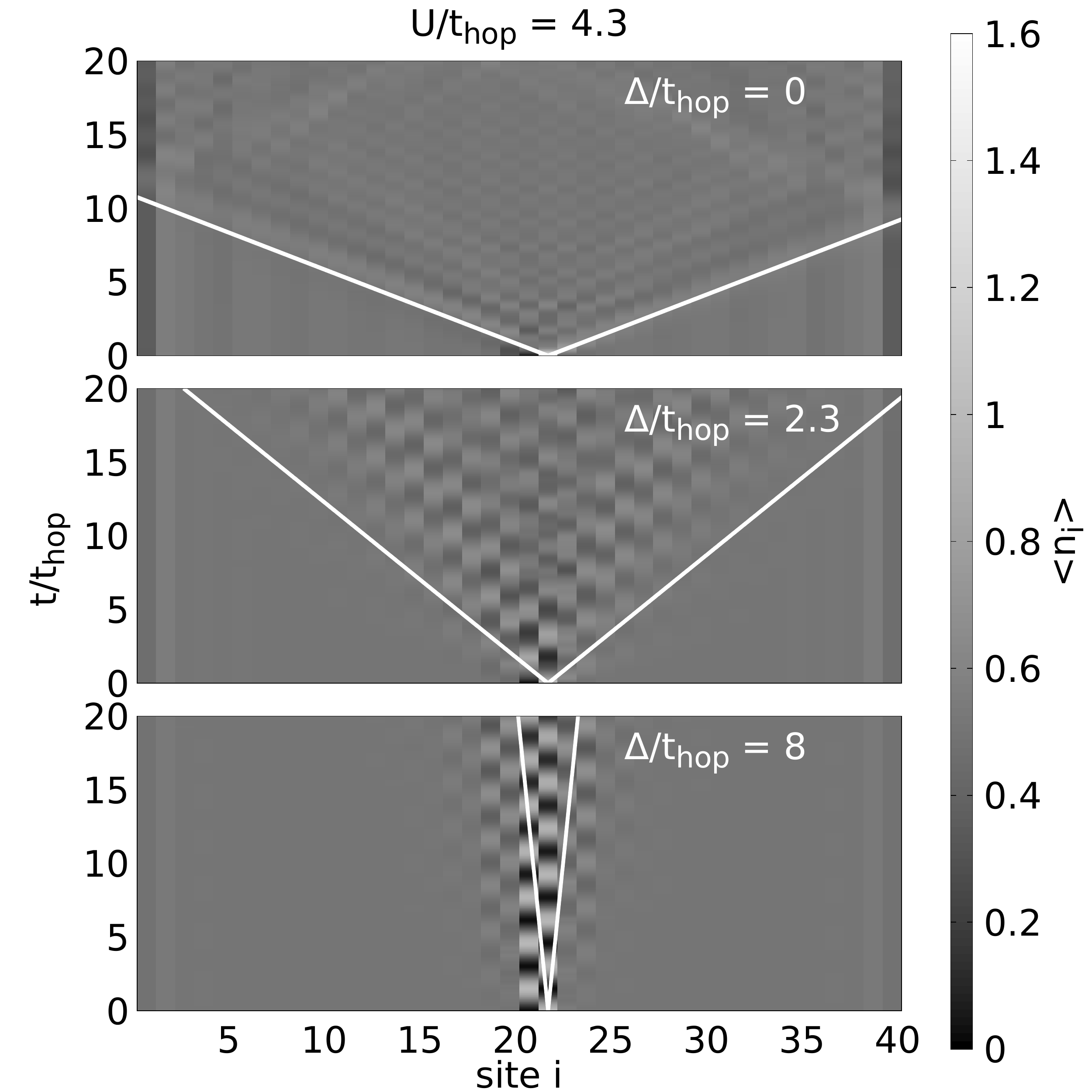}
	\caption
	{
		Time evolution of the local density $\langle \hat{n}_{i} \rangle$ following an excitation by applying operator Eq.~\eqref{eq:single_exciton} at the center of the system. 
		The panels show tDMRG results for different values of \(\Delta/\hopping\) for chains with $L=40$ lattice sites. 
		Left side: $U = 0$; right side: \(U/t_{hop}=4.3\). 
		The solid lines indicate the maximal group velocity of the excited electrons obtained from the noninteracting band structure Eq.~\eqref{equ:1p-band-structure}, assuming that one electron gets excited from the first to the second band. 
		The dotted and dashed lines indicate the phase velocity at the $k$-value with the maximal group velocity, as discussed in the text.
	}
	\label{fig:N_delta}
\end{figure*}

\begin{figure}[b]
	\includegraphics[scale=0.32]{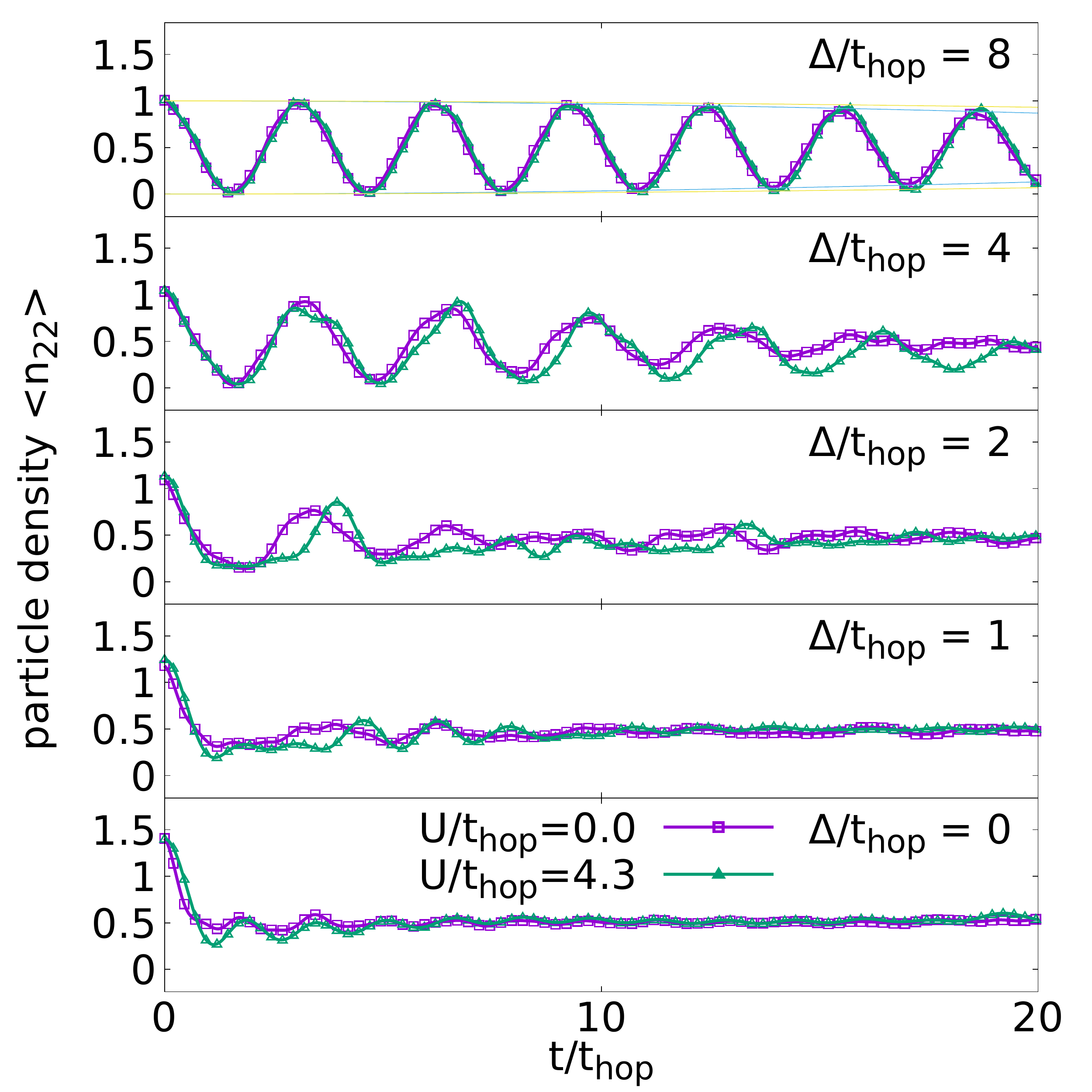}
	\caption
	{
		Time evolution of the local density $\langle \hat{n}_{R+1} \rangle$ obtained by tDMRG for a system with $L = 40$ lattice sites at the right site of the Zener polaron at which operator Eq.~\eqref{eq:single_exciton} was applied to, $R = L/2+1$. 
		Green: $U = 0$; purple: $U/\hopping = 4.3$. 
		The lines for \(\Delta/t_{hop}=8\) show a fit using a function of the form \(f(x)=\frac14\left(\cos(a x)+\cos(b x)\right)+\frac12\).
	}
	\label{fig:dmrg_delta_dependency}
\end{figure}

In \cref{fig:N_delta}, we show the tDMRG results for the time evolution of the local densities $\langle \hat{n}_R \rangle := \langle \hat{n}_{\uparrow, R} + \hat{n}_{\downarrow, R}\rangle$ following a photoexcitation.
This is modeled by applying operator Eq.~\eqref{eq:single_exciton} at the center of the system to the ground state obtained from a DMRG calculation, which induces a local dipole on the Zener polaron at the center of the system.
We display results for $\Delta/t_{hop} = 0,\,2.3,\ 8$ and compare the cases $U = 0$ (left panels) to the case $U/t_{hop} = 4.3$ (right panels). 
The case with $\Delta / \hopping = 2.3$ and $U / \hopping = 4.3$ corresponds to the values of Table~\ref{tab:parameters}. 

Let us start the discussion with the behavior at $\Delta = 0$.
In the ground state we observe Friedel-like density oscillations caused by the open boundary conditions used.\cite{Bedurftig:1998p457,White:2002p348}
They are typical for the Luttinger liquid phase\cite{giamarchi} realized in the Hubbard chain at this value of the filling.\cite{HubbBook} 
These Friedel-like oscillations are stable and do not change with time.

On top, we see that the local excitation created at the center of the system spreads through the lattice with constant maximum speed.
This light-cone behavior is captured by a Lieb-Robinson bound,\cite{LiebRobinson72} which states that in nonrelativistic quantum lattice systems with a short-ranged Hamiltonian information spreads with a finite maximal velocity.

In this case, for $U=0$ and $\Delta = 0$, the maximal group velocity allowed by the band structure Eq.~\eqref{equ:1p-band-structure} is the Fermi velocity $v_F = 2 \frac{\hopping \, a}{\hbar}$. 
In the units used ($a = \hopping = \hbar = 1$), this leads to a slope of $2$ in the lightcone, which is what is seen in Fig.~\ref{fig:N_delta} for $\Delta = U = 0$. 
For $U > 0$ and $\Delta = 0$, the velocity gets modified by the interaction, but as expected from Luttinger liquid theory,\cite{giamarchi} the system will always show ballistic motion of the excitation, i.e., it will propagate with a constant maximal velocity through the system. 

For finite values of $\Delta$ the Friedel-like oscillations disappear.
This is expected, since for any finite value of $\Delta$ a band gap is formed so that the Fermi surface vanishes, and with it the Luttinger liquid phase and the Friedel-like oscillations. 

By increasing the value of $\Delta / \hopping$, the velocity of the spread of the excitation is seen to decrease. 
For $U=0$ this is expected from the single-particle band structure Eq.~\eqref{equ:1p-band-structure}, in which the bands become flatter with increasing $\Delta / \hopping$, which also reduces the maximal group velocity.

For the times shown $t/\hopping \leq 20$ (corresponding to $\sim 23$~fs using the values of Table~\ref{tab:parameters}), for $\Delta/\hopping = 8$ the speed of the excitation is close to zero, since the group velocities obtained from the band structure are very small already (e.g., the maximal group velocity for an electron excited to the second band is $v \approx 0.08 \frac{\hopping a}{\hbar}$). 
At the site of the excitation, the dipole-like density oscillations become clearly weaker with time for $\Delta/\hopping = 2.3$ as the energy is transferred to the neighboring sites. 
For the largest Hund's splitting shown, $\Delta /\hopping = 8$, the dipole oscillations remain concentrated on the central site on the time scale shown.  

While for $\Delta = 0$ the noninteracting electrons move with the expected Fermi velocity $v_F = 2\frac{\hopping a}{\hbar}$, for the intermediate value $\Delta / \hopping = 2.3$ an interesting structure emerges, which is apparently caused by the presence of both dipole-like oscillations of the electron on the excited Zener polaron and the relatively small tunneling barrier between the polarons: 

When the electron reaches the boundary between two Zener polarons, it gets partially reflected, but can also partially tunnel to the next polaron. 
This happens again for both the transmitted as well as the reflected part of the electron when they reach the border to the next polaron, and so on.
The result is the intricate pattern seen in Fig.~\ref{fig:N_delta}, in which the excited electron seems to spread through the system in a ping-pong or billiard-like manner for $U=0$ and $\Delta/\hopping = 2.3$.
However, now a further interesting effect comes into play, which leads to linear structures with a slope substantially larger than the maximal group velocity allowed by the band structure.
This was discussed in Ref.~\onlinecite{preprintLorenzo} in the context of interacting Mott insulators: The spread of information through the lattice is governed by the Lieb-Robinson velocity, which here can be estimated as the maximal group velocity determined by $v_{g, \nu} = \partial \epsilon_\nu(k)/\partial k$, with the band $\nu$, to which the electron is excited to.
However, as described in Ref.~\onlinecite{preprintLorenzo}, within the light cone and in its vicinity it is possible to have linear structures with a slope corresponding to the maximal \textit{phase velocity} instead, which is determined via $v_{p, \nu} = \epsilon_\nu(k^*)/k^*$, where $k^*$ is the momentum, at which $v_{g, \nu}$ is maximal. 
The phase velocity can be substantially larger than the maximal group velocity. 
This corresponds to what is seen in Fig.~\ref{fig:N_delta} for $U=0$ and $\Delta/\hopping = 2.3$: The excitation causes linear structures, whose slope is in excellent agreement with the maximal phase velocity obtained from the band $\nu = 2$ in Eq.~\eqref{equ:1p-band-structure}. 
However, the structure is seen to be strong only as long as it is within or close to the light cone, which is obtained from the maximal group velocity determined from $\epsilon_2(k)$ in Eq.~\eqref{equ:1p-band-structure}.
As soon as they reach the border of the light cone, their amplitude decays quickly, so that they do not contribute to the spread of information through the lattice.

In the presence of repulsive $U$, it is an interesting question whether the ballistic transport will prevail, or if the interparticle scattering might change its speed, e.g., inhibiting transport by slowing down the spreading of the excitation, or enhancing transport by increasing its velocity.
Also, it is possible that transport at finite $U$ could change its nature from ballistic to diffusive, or that even at a relatively small value of $\Delta / \hopping$ the excitation might get trapped. 

The right side of Fig.~\ref{fig:N_delta} shows results for $U/\hopping = 4.3$.
For $\Delta =0$, as discussed above, the ballistic motion prevails, as expected for a Luttinger liquid.
At finite $\Delta / \hopping$, however, the behavior changes significantly when comparing to the corresponding $U=0$ cases: At $\Delta / \hopping = 2.3$, the ping-pong-like structure disappears and is replaced by a more diffuse looking behavior.
This is captured by the following scenario: Due to the rather strong interaction, the electron scatters as soon as it tunnels to the neighboring Zener polaron, since there the electron is of opposite spin, so that the Hubbard term comes into play.
This scattering induces on one hand a dipole oscillation also on this Zener polaron, and on the other hand a partial tunneling of the electron of opposite spin to the neighboring lattice site.
There, the mechanism repeats, and again a dipole-like oscillation also on this Zener polaron is excited, and partial tunneling of the electron with opposite spin direction to the further Zener polaron is induced, and so on. 
The resulting picture is a sequence of dipole oscillations formed on each Zener polaron, with an amplitude decreasing the further one moves away from the site of the excitation. 
This sequence of dipole oscillations seems to replace the ping-pong pattern observed at $U = 0$.
It is difficult to judge whether the motion of the original excitation through the system remains ballistic, or if it might change its nature.
However, the strongest features are deep inside the light cone prescribed by the group velocity of the noninteracting system and seem to move with a smaller velocity, or in a diffusive manner.

Also at large $\Delta / \hopping = 8$, the effect of a finite value of $U$ is significant: While at $U = 0$, on the time scales shown, there was essentially no spread of the excitation to the neighboring sites, now the dynamics is clearly composed of the dipole oscillation on the excited dimer, plus additional dipole oscillations on the close lying neighboring dimers.
Again, it is difficult to conclude whether transport might be diffusive or ballistic.
We leave this interesting aspect for future research. 

We complement this discussion by considering the time evolution of the local density $\langle \hat n_R \rangle$ on the excited dimer in more detail. 
In Fig.~\ref{fig:dmrg_delta_dependency}, we show our tDMRG results at $U = 0$ and $4.3$ for the different values of $\Delta / \hopping$ indicated there.  
In contrast to the different behavior seen in Fig.~\ref{fig:N_delta} when comparing the results for $U=0$ to the ones for $U / \hopping = 4.3$, in all cases shown and on the time scale displayed, the time evolution on the site of the excitation is qualitatively similar with and without interaction.  
On the time scale shown, three different types of behavior seem to exist: 
For $\Delta/ \hopping = 8$ the value of the local density shows a coherent oscillation for all times shown $t / \hopping \leq 20$ (corresponding to $\approx 23$~fs using the values of Table~\ref{tab:parameters}). 
The amplitude of this oscillation decays only slowly. 
As can be seen in Fig.~\ref{fig:N_delta}, the reason for this are the dipole oscillations on the Zener polaron where the excitation was created, which are present for both values of $U/ \hopping$. 
As the group velocity for the excitation moving away from this place is so small in this case, the dipole oscillations decay only slowly. 
For the local density, the effect of $U$ is to weakly dampen its oscillation.  

In the other extreme case displayed at $\Delta = 0$, one sees that the coherent oscillation of the local density is completely suppressed, and the value of the local density drops very quickly to the equilibrium value $0.5$ and then shows only tiny oscillations around this value. 
This drop happens on a time scale $t /\hopping < 10 $, corresponding to $\sim11$~fs using the parameters of Table~\ref{tab:parameters}. 
The reason for this is that the excitation moves freely through the system, as discussed above, so that at the site of the excitation the local density relaxes quickly to the equilibrium value, up to the small oscillations seen in Fig.~\ref{fig:dmrg_delta_dependency}. 
This is also true at finite $U$, where the system is in a Luttinger liquid phase.\cite{giamarchi} 

\begin{figure}[t]
	\includegraphics[width=0.48\textwidth]{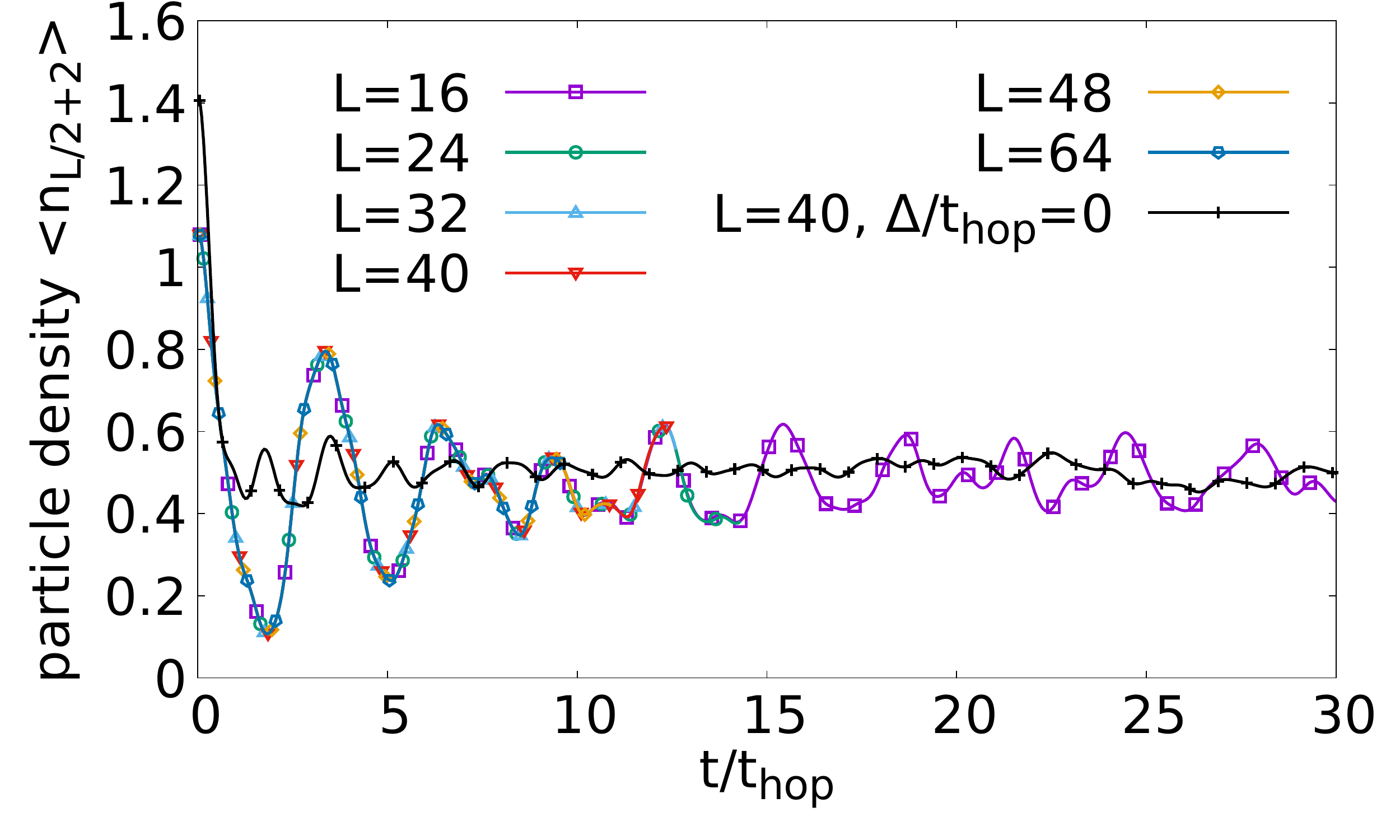}
	\caption
	{
		Time evolution of the local density $\langle \hat{n}_{R+1} \rangle$ after the operator Eq.~\eqref{eq:single_exciton} was applied to $R=L/2+1$ for $\Delta / \hopping = 2.3$ and $U/ \hopping = 4.3$, which is close to the parameters of \cref{tab:parameters}.
		The plot compares tDMRG results for systems with $L = 16$ (violet boxes), $L = 24$ (green circles), $L = 32$ (blue up-pointing triangles), $L = 40$ (red down-pointing triangles), $L = 48$ (yellow diamonds), and $L = 64$ (dark blue pentagons). 
		The results displayed are obtained with MPS matrix dimension $\chi_{\text{MPS}}=5000$.
		Additionally the time evolution for \(\Delta=0\) and \(U/\hopping=4.3\) for a system with \(L=40\) and \(\chi_{\text{MPS}}=500\) is plotted (black pluses).
	}
	\label{fig:dmrg_n_exc_vs_time}
\end{figure}

For intermediate values of $\Delta / \hopping$, the time evolution of the local density on this time scale $\lesssim 30$fs reflects both aspects:
At short times, coherent oscillations are seen, which are indicative for the dipole oscillation of the excited electron, whereas at later times the local density relaxes to its equilibrium value of $0.5$, since the excitation then is spreading through the system. 
Interestingly, the amplitudes of the oscillations around the equilibrium value are larger than for $\Delta = 0$ and do not depend on the system size, as can be seen in Fig.~\ref{fig:dmrg_n_exc_vs_time}, so that finite size effects seem to be excluded as cause for this behavior.

\subsection{Time evolution of the electronic momentum distribution function}
\label{sec:mdfs}

\begin{figure}[b]
	\includegraphics[width=0.48\textwidth]{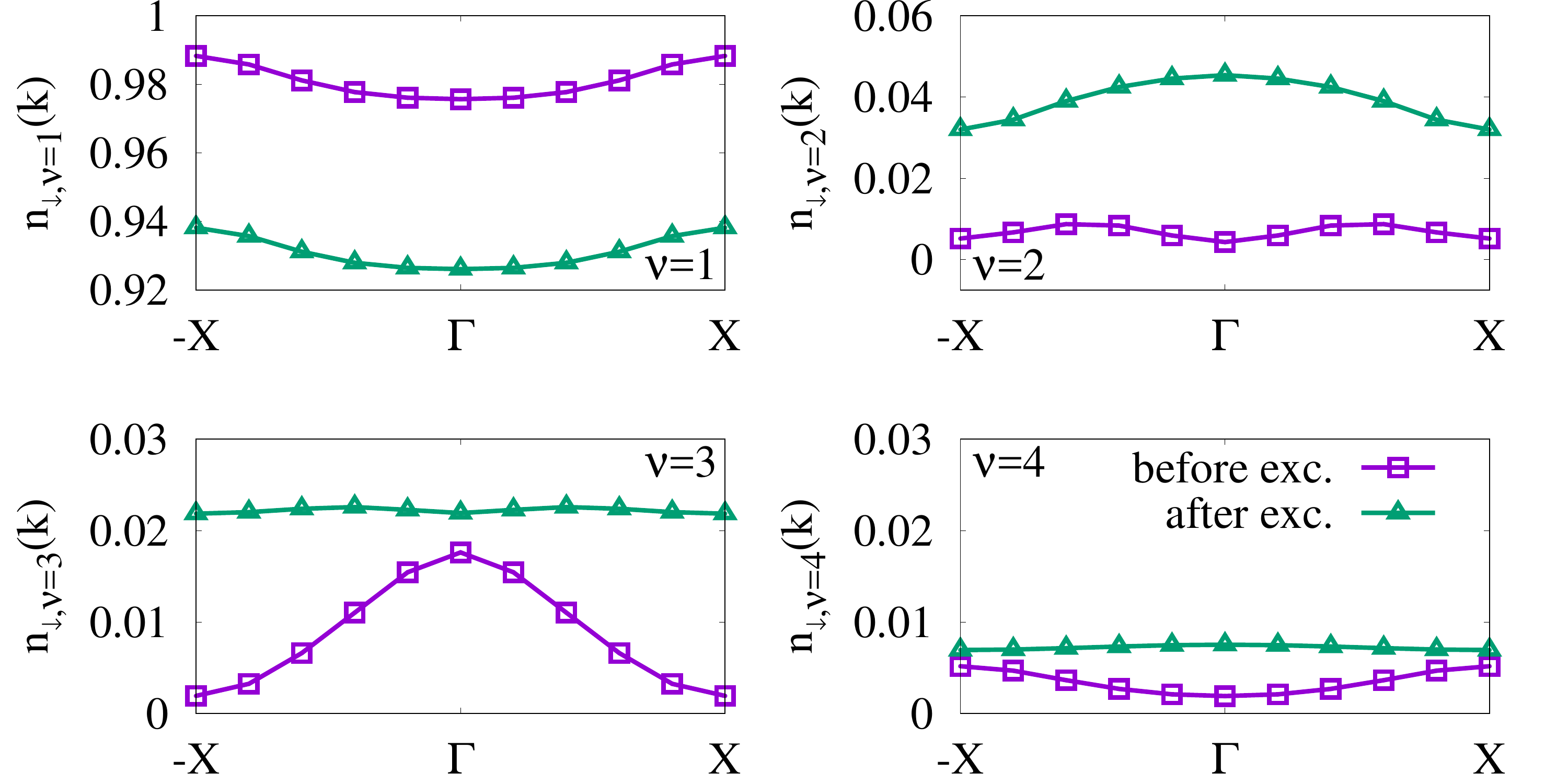}
	\caption
	{
		Momentum distribution for a system with $L=40$, $\Delta/\hopping = 2.3$, and $U/\hopping = 4.3$ before (magenta) and just after (green) the photoexcitation by applying operator Eq.~\eqref{eq:single_exciton} at the center of the system as obtained by the DMRG.
	}
	\label{fig:N_el_t0_Delta2}
\end{figure}

In this section, we present the time evolution of the momentum distribution at short times using the tDMRG, from which we obtain the time evolution of the electronic one-particle reduced density matrix 
\myEquBegin
\varrho_{\sigma,i,j}(t) = \langle\cd_{\sigma,i} \cw_{\sigma,j} \rangle (t)\; .
\myEquEnd
The time evolution of the momentum distribution is obtained by Fourier-transforming the one-particle reduced density matrix by projecting onto the four bands of the noninteracting system.
The momentum distribution of each band $\nu\in \mathds B = \{1, 2, 3, 4\}$ is then obtained by the corresponding transformation of the creation and annihilation operators (see \cref{app:LBE}), leading to
\myEquBegin
\label{eq:n_el}
\nel_{\sigma,\nu}(k,t) = \! \sum_{\mathclap{\begin{smallmatrix}\ell,\ell' \in \mathds Z \\ j,j' \in \mathds B \end{smallmatrix}}} e^{i2\pi k(\ell-\ell')} \, T^\ast_{\sigma\nu j} (k) \, \varrho_{\sigma,4\ell+j,4\ell'+j'}(t) \, T_{\sigma\nu j'}(k) \; ,
\myEquEnd
where the matrices $T_{\sigma\nu j}(k)$ are the unitary matrices holding the eigenvectors of the Hamiltonian of a single unit cell, as derived in detail in \cref{app:LBE}.
In \cref{fig:N_el_t0_Delta2}, we compare the momentum distribution of the ground state with the one obtained directly after the excitation. 
The system is excited by applying operator Eq.~\eqref{eq:single_exciton} to the central site of the system for $\Delta/\hopping = 2.3$ and $U/\hopping=4.3$. 

Note, that the excitation affects predominantly one spin direction, which is due to the spin polarization of the polaron on which the excitation takes place.
Let us first discuss the momentum distribution of the ground state.
Because we are at quarter filling, as expected, the first band $\nu=1$ is highest populated, and the population of the higher bands is negligibly small but finite since $U > 0$. 
Note that at $U / \hopping = 4.3$ the populations are slightly inverted, so that the momentum distribution at $k=0$ is somewhat smaller than at finite $k$. 
This is absent at $U=0$, as further discussed in \cref{app:U0}. 
At finite $U$, we associate this effect with the projection onto the noninteracting band structure.
It would be interesting to compare to the one-particle spectral function $A(k,\omega)$ at finite $U$, which can be measured in ARPES\cite{RevModPhys.75.473} experiments and which provides details of the band structure in the interacting case. 
As this exceeds the scope of this paper, we leave this aspect for future investigations.   
Here, we pursue a simpler path and consider the time evolution of the noninteracting bands and their populations as indicators for the strength of the scattering between the bands and for time scales emerging in the course of the time evolution.

The photoexcitation moves particles from the lowest band to the higher ones. 
As we model it as strongly localized in real space, the excitation here transfers all possible momenta in contrast to light, for which the momentum transfer is negligible. 
For $\Delta/\hopping=2.3$, the second and third band get a higher population, whereas the one of the fourth band remains very small. 
For the largest value of the Hund's splitting, $\Delta/\hopping = 8$ treated in the previous section, the most affected band is the second one; the population of the two highest bands remains very small.
Hence, the lowest band $\nu=1$ is highest populated in the ground state and remains highest populated also after the excitation in all cases treated.

\begin{figure}[t]
	\includegraphics[width=0.48\textwidth]{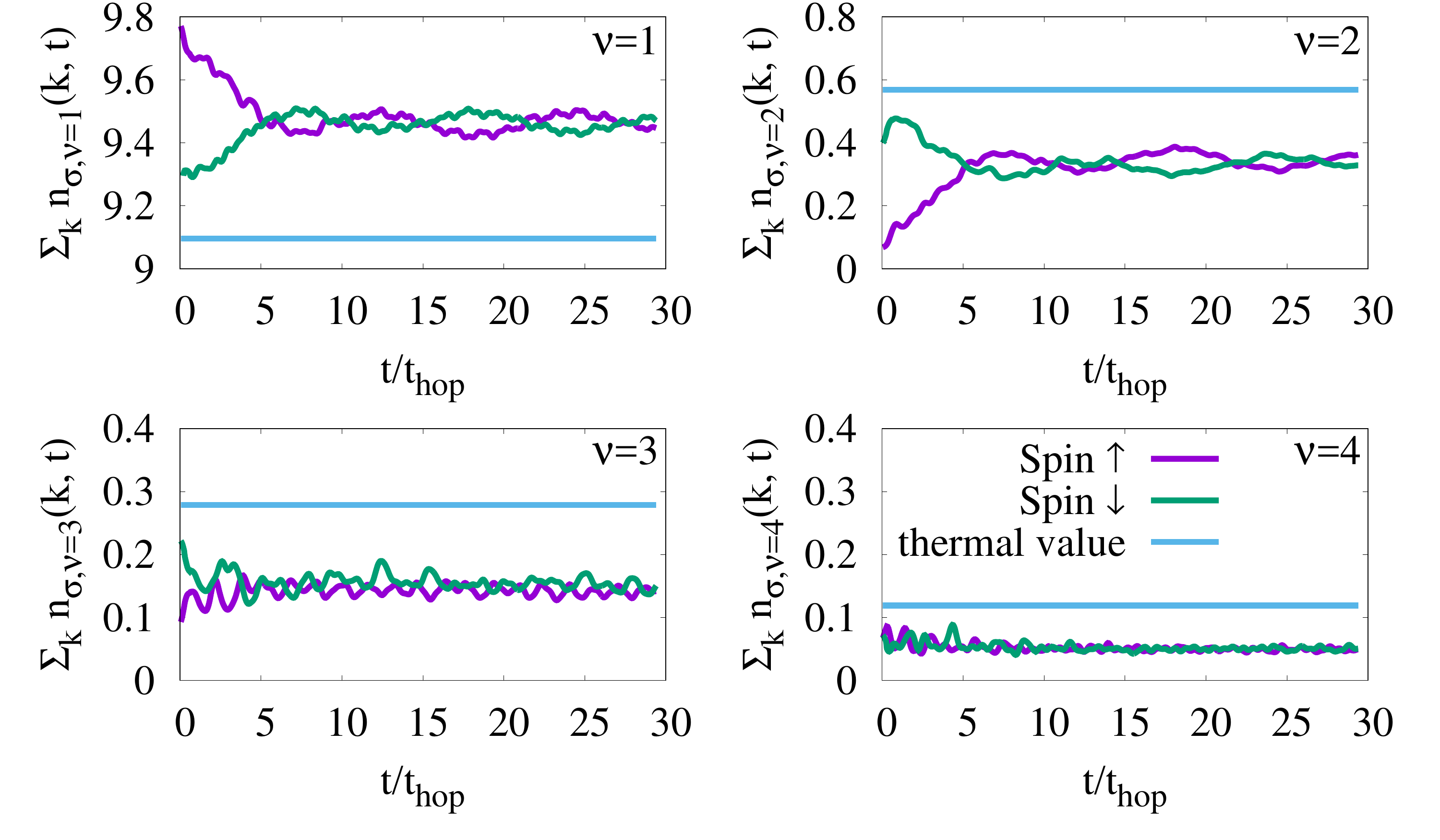}
	\caption
	{
		Time evolution of the population $\sum_k n_{\sigma,\nu}^{\text{el}}(k, t)$ of each band following the photoexcitation for $\Delta/\hopping =2.3$ and $U/\hopping=4.3$ as obtained by the tDMRG. 
		Additionally the population of the thermal state is given by the horizontal lines.
	}
	\label{fig:N_el_spinup_Delta2_single}
\end{figure}

\begin{figure}[b]
	\includegraphics[width=0.48\textwidth]{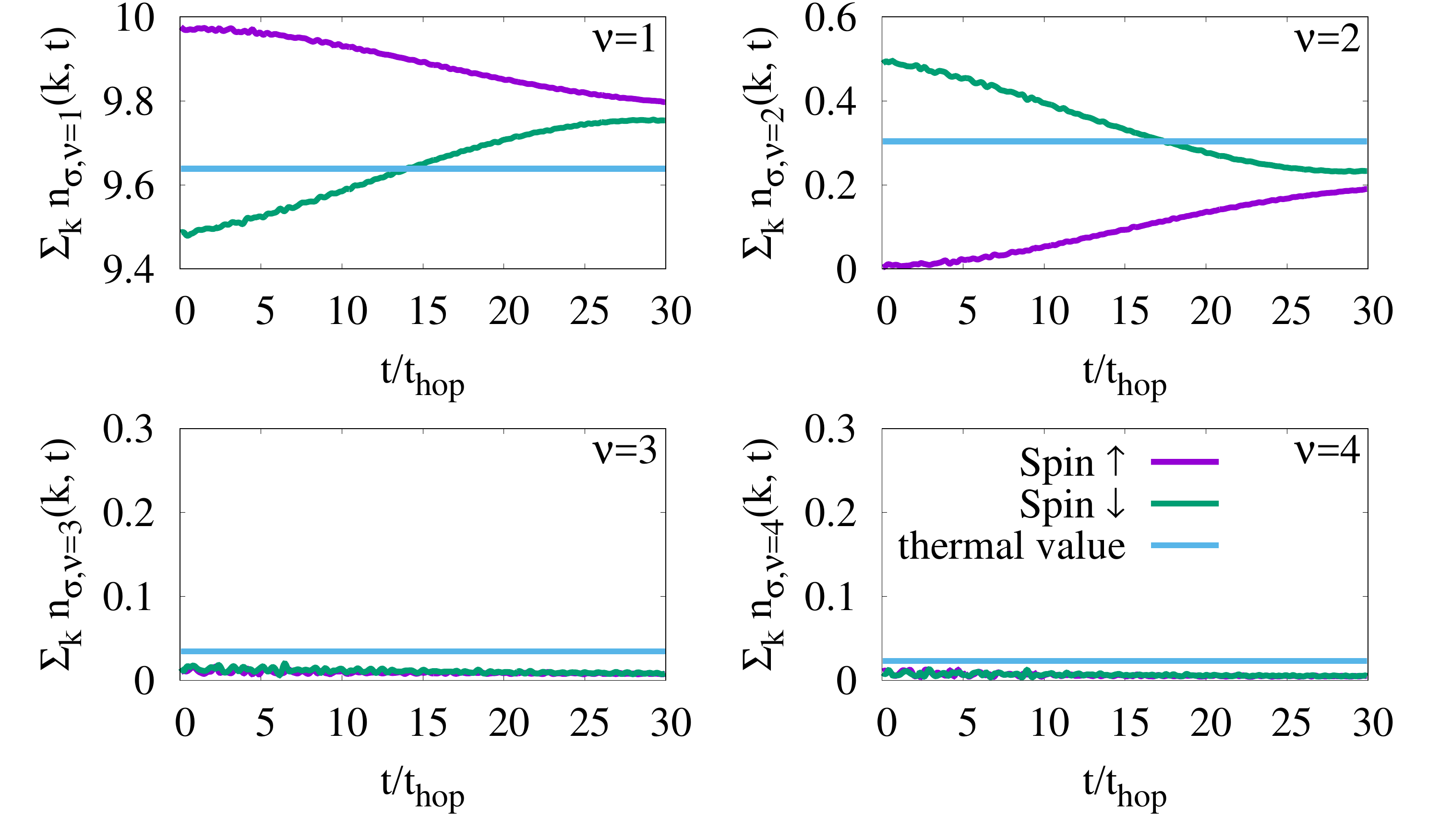}
	\caption
	{
		Time evolution of the population $\sum_k n_{\sigma,\nu}^{\text{el}}(k, t)$ of each band following the photoexcitation as in Fig.~\ref{fig:N_el_spinup_Delta2_single}, but for $\Delta/\hopping =8$, as obtained by the tDMRG.
		Additionally the population of the thermal state is given by the horizontal lines.
	}
	\label{fig:N_el_spinup_Delta8_single}
\end{figure}

Due to the finite value of $U / \hopping$, we expect the electrons to scatter so that the population of the four bands changes in time. 
In \cref{fig:N_el_spinup_Delta2_single,fig:N_el_spinup_Delta8_single}, we show the time evolution of the populations of both spin directions for each of the four bands $\sum_k n_{\sigma,\nu}^{\text{el}}(k, t)$ for $\Delta/\hopping = 2.3$ and $\Delta/\hopping = 8$, respectively.
Clearly, scattering between the bands takes place.
In contrast to the time evolution of the local densities treated in the previous section, the band populations in Figs.~\ref{fig:N_el_spinup_Delta2_single} and~\ref{fig:N_el_spinup_Delta8_single} are indicative for bulk behavior and hence are better suitable for identifying time scales, on which the excitation evolves. 
As displayed in Fig.~\ref{fig:N_el_spinup_Delta2_single} for $\Delta/\hopping = 2.3$, the populations of the  first and second band seem to relax to a stationary value of $\sim 9.45$ and $\sim 0.35$ on a time scale of $\sim 5 \hopping$ (corresponding to $\sim 6$fs using the parameters of Table~\ref{tab:parameters}).
The populations of both spin directions relax to the same value and afterwards show rather small oscillations around these values.
Similar behavior is also seen in the third and fourth band.
For $\Delta/\hopping = 8$, instead, relaxation happens only at a time $t/ \hopping> 30$. 
The first two bands seem to reach a population of $\sim 9.8$ and $\sim 0.2$, respectively.
The third and fourth band have very small populations. 
The population of both spin directions seems to relax to the same value, even though at $t=0$ they significantly differ. 

As seen in Figs.~\ref{fig:N_el_spinup_Delta2_single} and~\ref{fig:N_el_spinup_Delta8_single} the spin moment inside the bands seems to relax on a short time scale $\lesssim 50$~fs.
These results indicate that the relaxation time increases with the value of $\Delta/ \hopping$.
However, it is still possible that further aspects can become important for the lifetimes of the excitations. 
The question arises, if one can make a quantitative prediction for the lifetime of the excitation in the presence of $U$ and $\Delta$ also in cases, which are not amenable to the tDMRG.
As much longer times are barely accessible to the tDMRG, we therefore now turn over to the LBE treatment, which is suitable to extract lifetimes of the excitations.

\section{Quasiparticle relaxation}
\label{sec:relax}

In this section, we  use the numerically exact results for the time evolution obtained by MPS to estimate the quasiparticle content needed for a quantum Boltzmann equation (BE). 
The BE will then provide us with information about the long-time behavior after the excitation, which is inaccessible using the tDMRG because of the fast growth of entanglement\cite{Amico:2008en} with time.

\subsection{Calculation of the quasiparticle momentum distribution from the tDMRG results}
\label{sec:LBE-methods}

The applicability of the BE for our one dimensional model is justified by the same reasoning as in Refs.~\onlinecite{Spohn2012,Spohn2013,Biebl2016}, see also Refs.~\onlinecite{Erdoes2004,Spohn2009}: 
The quasifree property of the system persists up to time scales $\propto U^{-2}$ and intervening scattering processes with a rate $\propto U^2$ allow one to use the fermionic BE on all time scales. 

One important point to realize is that within Fermi liquid theory, the BE requires the quasiparticle distribution function as input and not the distribution function for the electrons themselves \cite{Kamenev2009} (notice that otherwise the zero temperature ground state of an interacting Fermi liquid would not be a fixed point of the BE). 
So as a first step, we need to find this quasiparticle momentum distribution from the tDMRG results.

For the sake of simplicity we suppress the band and the spin index in the following. 
The equilibrium distribution function of the electrons $n^{\rm el}(k)$ is defined via Eq.~\eqref{eq:n_el}. 
The quasiparticle distribution function $n^{\rm qp}(k)$ is the Fermi-Dirac distribution for the noninteracting band structure Eq.~\eqref{equ:1p-band-structure}, here at $T=0$.
For a Fermi liquid at zero temperature, the relation to $n^{\rm el}(k)$ in the vicinity of the Fermi surface is then:
\begin{equation}
\lim_{k\rightarrow k_F} \left\{ n^{\rm el}(k)-\frac{1}{2}-Z\,\left(n^{\rm qp}(k)-\frac{1}{2}\right)\right\}=0 \, .
\end{equation}
If we use this relation away from the Fermi surface, it defines a $k$-dependent quasiparticle residue $Z(k)$, which describes the spectral weight of the pole in the one-particle Green’s function for momentum~$k$ 
\begin{equation}
\label{eq_nel_nqp}
n^{\rm el}(k)-\frac{1}{2}=Z(k)\,\left(n^{\rm qp}(k)-\frac{1}{2}\right),
\end{equation}
which gives us the means to determine $Z(k)$ from the equilibrium distribution
\begin{equation}
Z(k)=\frac{n^{\rm el}(k)-\frac{1}{2}}{n^{\rm qp}(k)-\frac{1}{2}} \ .
\end{equation}
We now apply (\ref{eq_nel_nqp}) to the nonequilibrium situation as well:
\begin{eqnarray}
\label{eq_nelt_nqpt}
n^{\rm el}(k,t)-\frac{1}{2}&=&Z(k)\,\left(n^{\rm qp}(k,t)-\frac{1}{2}\right) \\
\Rightarrow\quad n^{\rm qp}(k,t)&=&\frac{1}{2}+\frac{1}{Z(k)}\left(n^{\rm el}(k,t)-\frac{1}{2}\right) \\
&=& \frac{1}{2} + \frac{n^{\rm qp}(k)-\frac{1}{2}}{n^{\rm el}(k)-\frac{1}{2}}\left( n^{\rm el}(k,t)-\frac{1}{2}\right). \ 
\end{eqnarray}
This yields the desired relation between the distribution function of the electrons $n^{\rm el}(k,t)$ measured by tDMRG and the quasiparticle distribution function of the quasiparticles $n^{\rm qp}(k,t)$ as input for the BE. 
The justification for the step from (\ref{eq_nel_nqp}) to (\ref{eq_nelt_nqpt}) comes from the continuous unitary transformation approach as used in Ref.~\onlinecite{Essler2014}: 
$Z(k)$ describes the spectral weight of the electron with quasimomentum~$k$ which propagates coherently between scattering processes described by the BE.  
This reasoning is a good approximation as verified by comparison with numerically exact results in Ref.~\onlinecite{Essler2014}. 
It is important to note that Eq.~\eqref{eq_nelt_nqpt} is only applicable after the short-time regime in which quasiparticles form.

\subsubsection{DMRG results for the Momentum-distribution function of the quasiparticles}
\label{sec:MDF_qp}

\begin{figure}
	\includegraphics[width=0.48\textwidth]{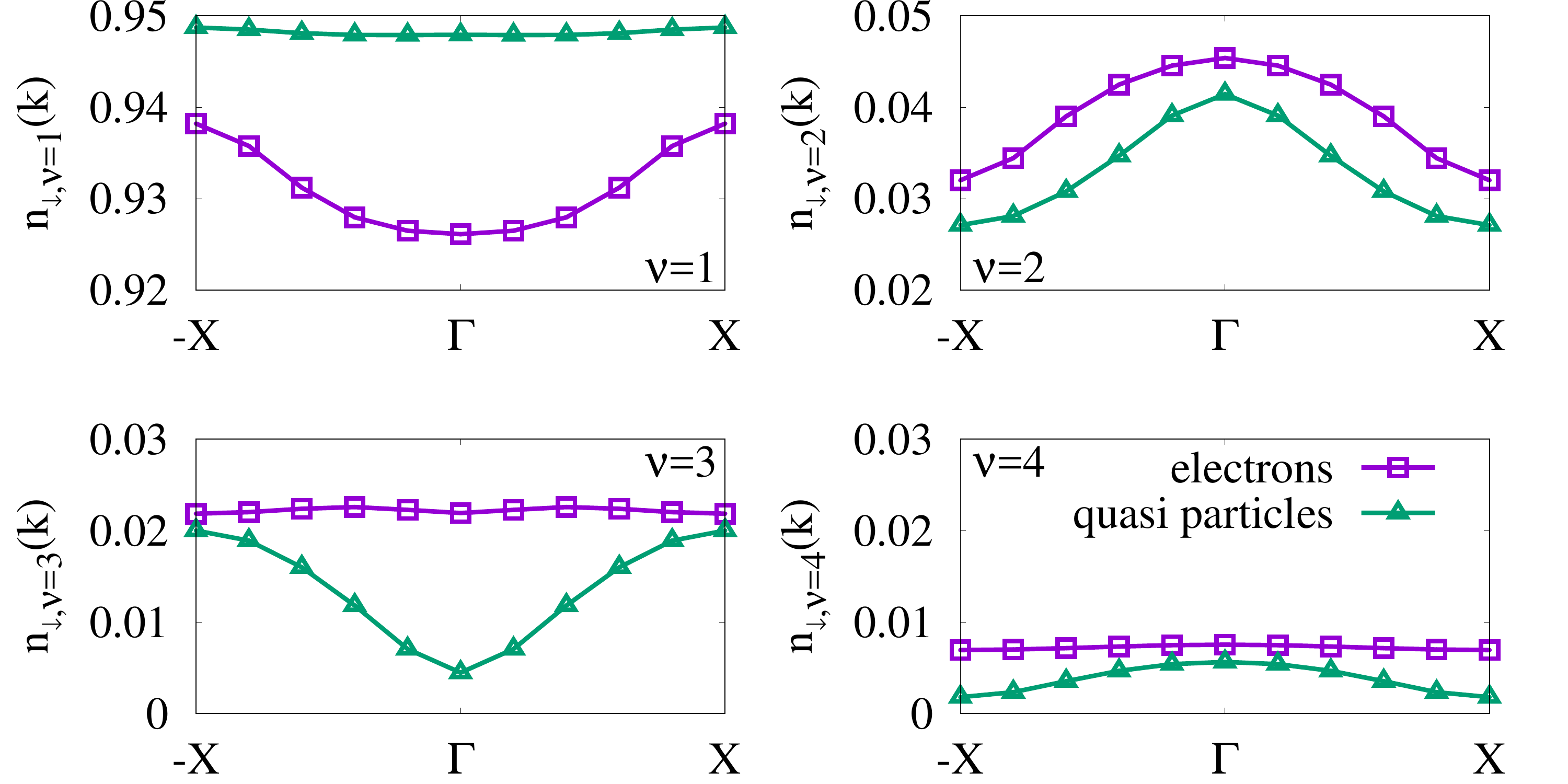}
	\caption
	{
		Comparison of the electronic and the quasiparticle momentum distributions obtained with DMRG via Eq.~\eqref{eq_nelt_nqpt} for a system with $L=40$, $\Delta/\hopping = 2.3$, and $U/\hopping = 4.3$ just after the photoexcitation by applying operator Eq.~\eqref{eq:single_exciton} at the center of the system.
	}
	\label{fig:N_qp_spinup_t0_Delta2}
\end{figure}

In \cref{fig:N_qp_spinup_t0_Delta2}, we show the quasiparticle distribution obtained from Eq.~\eqref{eq_nelt_nqpt} at the beginning of the time evolution after applying operator Eq.~\eqref{eq:single_exciton} at the center of the system. 
Although we expect the quasiparticle picture to be better applicable at later times, it is nevertheless instructive to compare these electronic and quasiparticle distributions to each other at $t=0$. 
According to Eq.~\eqref{eq_nelt_nqpt}, we expect a renormalization by a  $k$-dependent quasiparticle residue $Z(k)$, which is, however, constant in time.
Thus, the time evolution of the quasiparticle momentum distribution will be similar to the one of the electrons, if the renormalization is not too strong. 
For both values of $\Delta$ shown, the renormalization of the electronic momentum distribution is very small.
This is an important finding and justifies the following treatment with LBE. 

\subsection{Linearized multi-band Boltzmann equation for long-time relaxation}
\label{sec:LBE-model}

Based on the effective model from~\cref{sec:1D-model-valence-el}, Eq.~\eqref{equ: eff model Hamiltonian}, we can investigate the relaxation of the electrons due to electron-electron interactions by means of a quantum BE. 
We use it in a similar manner as Biebl and Kehrein in Ref.~\onlinecite{Biebl2016}, who investigated the thermalization rates of a Hubbard model with next-nearest-neighbor hopping. 
Furthermore, we perform a linearization of the BE to determine the relaxation rates.

To investigate the relaxation of the \QMD $\nqp_{\sigma,\nb}(k,t)$, we use the multiband \BE
\myEquBegin
\indfct[\hspace{-1pt}]{\dotnqp}{\sigma,\nu}{k,t} = \indfct[\hspace{-1pt}]{\Icoll[\nqp]}{\sigma,\nu}{k,t} \, ,
\label{eq:multibandBE}
\myEquEnd
with the collision term $\Icoll[\nqp]_{\sigma,\nu}(k,t)$.

With the BE, we can estimate arbitrary time scales, which can also be longer than the spin relaxation time. 
In the following we assume that the $\up$ particles have the same quasiparticle momentum distribution $\nqp_{\sigma,\nb}(k,t)$ as those with spin $\down$. 
Hence, the spin index $\sigma$ is not written explicitly any more, i.e.,~$ \nqp_{\nb}(k,t) = \nqp_{\up,\nb}(k,t) = \nqp_{\down,\nb}(k,t)$. 

The collision term is 
\begin{eqnarray}
\label{equ: multi-band-BE 0}
&&\mathcal{I}^{(\text{coll})}_{\nu_1}(k_1)
= 
\left(\frac{4a}{2\pi}\right)^2
\frac{\pi U^2}{\hbar \hopping}
 \int dk_2 dk_3 dk_4\sum_{\nu_2,\nu_3,\nu_4 \in \mathds B}
\nonumber\\
&&\times\PhiintnkII 
\sum_G
\delta(\dKX[\kvec] \! + \! G) \delta\bigl(\Domegank \bigr) 
\nonumber\\ 
&&\times \Bigl\{ \underbrace{\bigl[1 \smns \fctnkt{\nqp}{1}\bigr] 
\bigl[1 \smns \fctnkt{\nqp}{2}\bigr]  \fctnkt{\nqp}{3} \, \fctnkt{\nqp}{4}}_{\text{gain term}} 
\nonumber\\ 
&& \hphantom{\times \Bigl\{ }\underbrace{- \, \fctnkt{\nqp}{1} \, \fctnkt{\nqp}{2} \bigl[1\smns \fctnkt{\nqp}{3} \bigr] \bigl[1 \smns \fctnkt{\nqp}{4} \bigr]}_{\text{loss term}} \Bigr\} 
\,,
\nonumber\\
\end{eqnarray}
with $\vec \nu = \left(\nu_1 , \ldots, \nu_4 \right), \vec k = \left(k_1 , \ldots, k_4\right)$.
Furthermore, we define $\Phiintnk$ as the matrix element of the interaction taken at four momenta in four bands, see Eq.~\eqref{equ:LBE-Phi_k}.
The momentum is conserved by $\delta(\dKX)$ with $\dKX=k_1+k_2-k_3-k_4$ and the reciprocal lattice vector $G$.
It is important to note that the summation over $G\in\mathds{Z}$ is not an artificial addition but emergent from the derivation of $\Icoll$.
We express the energy conservation via $\Domegank:=\frac1\hopping\left(\epsilon_1+\epsilon_2-\epsilon_3-\epsilon_4\right)$ obtained from the one-particle band structure Eq.~\eqref{equ:1p-band-structure}.

The collision integral of a model describing an infinite lattice naturally allows for Umklapp processes. 
We account for this by integrating $k_4$ over a region larger than only the first Brillouin zone. 
We want to emphasize that this is an exact reformulation of the derived collision integral.

\def \betaRelaxTimeSplitting{1}

In order to investigate the long time relaxation, we linearize the \BE around the thermal distribution $\fdnk0 = 1/\{1+\exp [ \beta (\epsilon_{\nb}(k)-\mu) ]\}$. 
Here $\mu$ is the chemical potential and $\beta$ is the inverse final temperature. 
It can be determined via DMRG by setting the total energy of the system after the photoexcitation equal to the corresponding thermal expectation value, $\langle \hat{H} \rangle(t) = \frac{\rm Tr \left(\exp\left[-\beta \hat{H} \right] \hat{H}\right)}{\rm Tr \left(\exp\left[-\beta \hat{H} \right] \right)}$, for details see \cref{finite_temp_app}.

In order to perform the linearization, we define a perturbation $\phi_{\nb}(k,t)$ by\citep{Haug1996}
\myEquBegin
		\label{eq:perturbation}
		\nqp_{\nu}(k,t) = \frac{1}{ \ 1+\exp\{ \beta [\epsilon_{\nb}(k)-\mu] - \phi_{\nb}(k,t) \} \ } .
\myEquEnd
Note that at this point we could also use the numerical results for $\nqp_{\nu}(k,t)$. 
However, as discussed in Sec.~\ref{sec:MDF_qp}, the quasiparticle distribution obtained from the tDMRG is very similar to the electronic one.
The conceptually simplest approach is therefore to follow Ref.~\onlinecite{Biebl2016} and assume the quasiparticles to possess a distribution function as in Eq.~\eqref{eq:perturbation}. 
In future investigations, this can be refined by directly using the numerical results at large-enough times.

Equation~\eqref{eq:perturbation} leads to the linearized \BE
\myEquBegin
		\dotphinkt0 = \Lc[\phi]_\nu(k,t) .
\myEquEnd
The operator $\Lc$ acts on the perturbation $\phi$ and returns the change of the perturbation:
\myEquBegin
		\label{equ:LBE-linear-operator}
		\Lc[\phi]_{\nu_1} &= 
\left(\frac{4a}{2\pi}\right)^2\frac{ \pi U^2}{\hbar t_{hop}}
\int\! dk_2dk_3dk_4 \sum_{\nu_2,\nu_3,\nu_4\in\Sb}
\\
&\times
\PhiintnkII \ddKX \, \delta(\dKX[\kvec] \! + \! G) \\ 
				& \quad \times \frac{\bigl[ 1\!-\!\fctnkt{\fd}{2}\bigr] \fctnkt{\fd}{3} \fctnkt{\fd}{4}}{\fctnkt{\fd}{1}} \\ 
				& \quad \times \bigl[ \fctnkt{\phi}{1} \spls \fctnkt{\phi}{2} \smns \fctnkt{\phi}{3} \smns \fctnkt{\phi}{4} \bigr] .
\myEquEnd
$\Lc$ is hermitian in the scalar product
\myEquBegin
\label{equ: fscalprod sec1}
\fscalprod \phi \psi :=& 4a \int\frac{dk}{2\pi}  \, \sum_{\mathclap{\nu\in\Sb}} \phink0 \ \fdnk0 \bigl[ 1-\fdnk0 \bigr] \ \psink0 ,
\myEquEnd
which induces the norm $\fnorm{\phi}=\sqrt{\fscalprod{\phi}{\phi}}$. 
Therefore, we can represent the perturbation $\phinkt0$ by the eigenfunctions $\EFjnk j0$ and eigenvalues $\lamX_j$ of $\Lc$:
\myEquBegin
\phinkt0 = \sum_j \AjO j \ e^{- \lamX_j t} \  \EFjnk j0 .
\myEquEnd
The amplitudes $\AjO j=\fscalprod{\phinkt[0]0}{\EFj j}/\fnorm{\EFj j}^2$ are the overlaps of the respective eigenfunction and the initial perturbation $\phinkt[0]0$ at time $t=0$. 
The eigenvalues of $\Lc$ are the relaxation rates of the corresponding contribution $\AjO j$. 
One can proof that $\Lc$ is positive definite, i.e., its eigenvalues are non-negative. 
As long as the eigenvalues are positive, the factor $e^{- \lamX_j t}$ leads to the decay of the corresponding contribution of the perturbation $\phinkt0$. 
For $\lamX_j=0$, the respective part of the perturbation $\phinkt0$ does not decay.

There are two eigenvalues that are zero for any choice of the model parameters:
$\EFjnk10=\const$, and $\EFjnk20=\epsilon_{\nu}(k)$. 
For both, the factor $[ \phi_1 + \phi_2 - \phi_3 - \phi_4 ]$  in \cref{equ:LBE-linear-operator} vanishes. 
They correspond to conservation of particles and conservation of energy, respectively.

\subsection{Relaxation rates from Linearized Boltzmann Equations}
\label{sec:relaxationratesLBE}

\putFigureRelaxationRates{\label{fig: relaxation rates} 
	Relaxation rates as obtained from the linearized Boltzmann equation approach for $\DelX=1,2,4$ from left to right.  
	The eigenvalues $\lambda_n$ of $\Lc$ are sorted by their magnitude, and we plot the lowest ones ($n=1,2,3,...$) as a function of inverse temperature $\beta$.
	In all of the plots, the full lines are calculations with the lowest two bands, the dashed lines indicate a calculation with the three lowest bands, and the dotted lines denote a computation with all four bands.
	The eigenvalues $n=1,2,3$ are zero within numerical precision. 
	Note that, for the $3$- and $4$-band calculations, there is an additional zero eigenvalue, which we omit for comparability. 
	As two examples for relaxation times, we consider $u=U/\hopping=4.3$ at room temperature $\beta = t_{hop}/(\kB\cdot\unit[300]{K})\approx 23$ and at the temperature after excitation, which we estimate in \cref{finite_temp_app}, $\beta = 1.4$. 
	With $\Delta = 2.3$, $\hopping \approx 0.585$eV, and $t_0:=\hbar/\pi \hopping \approx \unit[0.12]{fs}$ the smallest relaxation time is $1/\lamX_4 = 10^{9} t_0/u^2 \approx \unit[6.5]{ns}$ for room temperature and $1/\lamX_4 = 10^{3} t_0/u^2 \approx \unit[6.5]{fs}$ for the temperature after the excitation treated in \cref{sec:photo}.
}

The relaxation rates are found by diagonalizing the dimensionless linear operator $t_0 (t_{hop}/U)^{2} \Lc[\phi]$ with the time scale $t_0=\hbar/2\pi \hopping \approx \unit[0.18]{fs}$.  
The result of our numerical evaluation is shown in \cref{fig: relaxation rates}.
There, we plot the results for a $2$-band, a $3$-band, and a $4$-band calculation, in which we always use the lowest bands possible. 
For final inverse temperature above $\beta\approx\betaRelaxTimeSplitting$, the calculations reveal the same relaxation rates.
This means that the upper bands are not involved in the relaxation for low temperatures.
Moreover, for low temperature, the relaxation rates decay exponentially in $\beta$.

For inverse temperatures larger than $\beta=30$, the lowest eigenvalues are zero within numerical precision.
Thus, for low temperatures, the corresponding contributions to the perturbation $\phinkt0$ become frozen.

The eigenvalues $n=1,2,3$ depicted in \cref{fig: relaxation rates} are numerically zero for every $\beta$.
For the $3$-band and $4$-band calculations, there is an additional zero eigenvalue $\lambda_+$, which we exclude from \cref{fig: relaxation rates} for a better comparison of the other eigenvalues.
We can explain these zero relaxation rates analytically.
They correspond to the eigenfunctions
\myEquBegin
\bigl( \EFjnk 10 \bigr)_{\nu=1,..,4} &= \bigl( 1,1,1,1 \bigr) \,\forall k, \\ 
\bigl( \EFjnk 20 \bigr)_{\nu=1,..,4} &= \bigl( \epsilon_1(k), \epsilon_2(k), \epsilon_3(k), \epsilon_4(k)\bigr) \,\forall k,
\\ \bigl( \EFjnk 30 \bigr)_{\nu=1,..,4} &= \bigl( 1,0,1,0 \bigr) \,\forall k,
\\ \bigl( \EFjnk +0 \bigr)_{\nu=1,..,4} &= \bigl( 1,1,0,0 \bigr) \,\forall k.
\label{equ: LBE conserved}
\myEquEnd
Because $\Lc$ is linear, all combinations of these eigenfunctions are conserved.
They correspond to quantum-mechanical state-space operators of the form $\SSop[\psi] = \int dk \, \psink0 \, \nqp_\nb(k)$:
\myEquBegin
\SSop[\EFj1] &= \N \ \  \text{ (total number of particles)},
\\ \SSop[\EFj2] &= \Ham + \order(U) \ \ \text{ (total energy)},
\\ \SSop[\EFj3] &= \N_1 + \N_3,
\\ \SSop[\EFj+] &= \N_1 + \N_2,
\myEquEnd
with the band number operators $\N_\nu = \int dk \, \nqp_\nb(k)$.
Hence, a contribution of type $\EFjnk10$ leads to a change of the total number of particles $\langle \N \rangle$. 
Likewise, a contribution of type  $\EFjnk20$ changes the energy density.
However, by construction of (\ref{eq:perturbation}) the initial perturbation $\phi_\nu(k,0)$ has zero overlap with
$\EFjnk10$ and $\EFjnk20$ and we can ignore these two eigenfunctions with vanishing rates. 

The eigenfunction $\EFjnk30$ means that the number of particles in the first plus those in the third band cannot be changed during the relaxation process.
Similarly, $\EFjnk+0$ is related to the conservation of particles in the first plus the second band.
Obviously, this eigenfunction is the same as $\EFjnk10$, if we do not include the bands $3$ and $4$.
This is the reason for the additional zero eigenvalue $\lamX_+$ in the $3$- and $4$-band case.
A contribution from either of the eigenfunctions $\EFjnk30$ and $\EFjnk+0$ leads to at least two different chemical potentials in the long time limit.
The cause of this is the relatively large value of $\DelX$.
Therefore, the gaps between the bands are so large that some of the two-particle scattering processes are forbidden by energy conservation.
More specifically, an interband relaxation requires the energy to be picked up by other interband or multiple intraband excitations. 
The limitation of scattering processes to two-electron processes, as in our study, suppresses the possibility of interband relaxation at the expense of increasing the intraband temperature.
Higher-order scattering processes will eventually become important for low final temperature.

The relaxation rates depend very sensitively on the final temperature and thereby the energy density of the photoexcitation.
In the case treated in Sec.~\ref{sec:photo}, we treat systems with typically 40 lattice sites, in which one electron gets excited.
As discussed in more detail in \cref{finite_temp_app}, this corresponds to an energy density leading to a final inverse temperature $\beta \approx 1.4$. 
For such high temperatures, the relaxation rates obtained from the LBE lead to a relaxation time scale $\sim 5-100$~fs, depending in detail on the values of $\Delta/ \hopping$ and $U/\hopping$. 
This prediction can now be compared to the numerical tDMRG results. 
As can be seen in Figs.~\ref{fig:N_el_spinup_Delta2_single} and~\ref{fig:N_el_spinup_Delta8_single}, the tDMRG results indicate that band occupations of the first band seem to relax to expectation values, which agree with the thermal expectation values up to a few percent.
Particle number conservation then leads to a difference of the band occupations in the other bands of similar absolute magnitude. 
This discrepancy can be due to the choice of boundary conditions and finite size effects, so that the results seem to be in good agreement with the corresponding thermal state.
As the LBE predicts a relaxation to a thermal state on comparable time scales, we conclude that the LBE treatment has predictive power for estimating relaxation rates also in other cases.
For example, when choosing an energy density such that the final temperature is of the order of room temperature ($\beta \approx 23$), for $\Delta / \hopping = 2.3$ and $U/\hopping = 4.3$, we obtain life times 
of several nanoseconds.
We believe this estimate for time scales can be useful to guide future experiments on similar systems.

\section{Conclusions and summary}
\label{sec:conclusions}

We present a combined theoretical approach to treat typical aspects of the relaxation behavior of photoexcitations in correlated materials over a wide range of time scales.
Specifically, we combined tight-binding models, which describe the interplay of electrons, spins and phonons, with numerically exact tDMRG studies and kinetic calculations using the linearized quantum Boltzmann equation. 

In order to alleviate the difficulties related to higher dimensions, we performed our study on a hypothetical one-dimensional manganite. 
This limitation to a one-dimensional material simplifies the description in several points: It helps to visualize the complex polaron and spin orders, it permits the study of the initial relaxation processes using tDMRG, which works particularly well in one-dimensional systems, and finally, it simplifies the high-dimensional integrals required for the collision term in the linearized quantum Boltzmann equation.

The tight-binding calculations showed that a polaronic microstructure is realized.
This can be described in an effective way in terms of the aggregation of various types of polarons, such as electron, hole, Zener, and Jahn-Teller polarons. 
Hence, the low-energy scale of 1D manganites can be well described in terms of polarons as basic entities, their reactions and interactions. 
This description provides a blueprint on how to rationalize the complex orbital, polaron, and spin orders in real materials in higher dimensions. 
Furthermore, it is a promising route towards more coarse grained simulations of the relaxation dynamics on the very long time scale dominated by the polaronic order.

The subsequent calculations have been performed with the spin and lattice degrees of freedom frozen in. 
Consistent with the electronic structure obtained from the tight-binding model, the electrons experience a sequence of Zener polarons, which are Mn dimers. 
Each dimer has two ferromagnetically coupled Mn sites, while two Zener polarons are antiferromagnetically coupled with each other. 
The resulting Hubbard-like model is controlled by three parameters, the hopping $\hopping$ between two Mn sites, the Hund's splitting $\Delta$ between spin up and spin-down electrons, and the Coulomb interaction $U$ between the electrons.

The excitation and the short-time initial relaxation has been studied using tDMRG simulations. 
In the absence of an interaction, the excitation of a specific Zener polaron produces internal dipole oscillations, which can be described as an electron-hole pair. 
Since the excitation is local in real space, it is spread over the entire reciprocal unit cell in momentum space. 
Electrons and holes propagate with a velocity determined by the slopes of the band structure. 
An intricate pattern of dipole oscillations inside the light cone of the excitation emerges at intermediate values of $\Delta/\hopping$, with structures propagating with the phase velocity at the $k$ value of the maximal group velocity, rather than with the group velocity itself. 
The group velocity decreases with increasing Hund's splitting, as it effectively decouples Zener polarons.

The electron-electron interaction $U$ induces a coupling of the dipole oscillations with different spins. 
It thus is responsible for a very rapid relaxation of the magnetic moment of the individual bands. 
While the role of $U$ is secondary in equilibrium, it has a pronounced effect on the dynamics, which differs strongly from that of noninteracting electrons. 
The tDMRG results for the momentum distribution clearly show that scattering due to the electron interaction leads to a redistribution of the electrons on the four bands on a femtosecond time scale. This timescale increases with $\Delta$ for fixed $U$.

The estimate of the quasiparticle content from the electronic momentum distributions reveals that only a small renormalization is present, so that the time evolution of the quasiparticle momentum distribution is very similar to the one of free electrons.

The relaxation rates have been determined with the linearized quantum Boltzmann equation. 
For all values of Hund's splitting, the lifetimes scale as $\sim \hopping / U^2$. 
A large Coulomb interaction increases the rate of scattering and thus increases the relaxation rate. 
As shown in Fig.~\ref{fig: relaxation rates}, a strong polaron microstructure expressed by the Hund's splitting $\Delta$ leads to an enhanced lifetime of the excitations. 
This finding is in agreement with the tDMRG results. 

The thermalization rates in our PCMO model are always exponentially suppressed as a function of the inverse final temperature (this is different from Ref.~\onlinecite{Biebl2016}), which would be a clear experimental signature if initially the sample is at sufficiently low temperature. 
This exponential suppression is a specific consequence of the one-dimensional nature of our PCMO model. 
However, it should be noted that depending on the strength of the electron-phonon coupling this signature might be hidden by coupling to the phonon bath. 
Realistic modeling and inclusion of the phonon degrees of freedom is left to future work.

\begin{acknowledgments}
	We acknowledge fruitful discussions with Z. Lenar\v{c}i\v{c}, F. Heidrich-Meisner, E. Jeckelmann, B. Lenz, L. Cevolani, N. Abeling, M. Schmitt, and H.-G. Evertz.  
	Financial support from the Deutsche Forschungsgemeinschaft (DFG) through SFB/CRC1073 (projects B03 and C03) and Research Unit FOR 1807 (project P7) is gratefully acknowledged. 
	S.R.M. acknowledges hospitality of the Kavli Institute for Theoretical Physics (KITP), Santa Barbara, where part of this research was accomplished and  supported in part by the NSF under Grant No. NSF PHY11-25915. 
	We acknowledge numerous insightful discussions with Thomas Pruschke (deceased). 
	S.R. and P.B. thank Robert Schade for his help with the tight-binding code. 
	We thank Christian Jooss for numerous discussions on the manganites, and Michael Seibt and Tobias Meyer for discussions on ongoing experiments.
\end{acknowledgments}

\begin{appendix}
\section{Boltzmann equation  for electron relaxation}
\label{app:LBE}

Based on the effective model from~\cref{sec:1D-model-valence-el}, Eq.~\eqref{equ: eff model Hamiltonian}, we investigate the relaxation of the electrons due to electron-electron interactions by means of a \BE. 
We use it in a similar manner as Biebl and Kehrein,\cite{Biebl2016} who investigated the thermalization rates of a Hubbard model. 
Furthermore we perform a linearization of the \BE to find the relaxation rates.

\subsection*{Multi-band \BoltzmannEquation}

In this appendix, we describe the definitions used in setting up the \BE.
A first step towards calculating the \BE is finding the one-particle bands $\epsilon_\nb(k)$ as eigenvalues of the noninteracting Hamiltonian~$\HO$. 

We consider a unit cell with four Mn-sites. The noninteracting Hamiltonian for a system with N unit cells has the form
\begin{eqnarray}
	\hat{H}_0=
	\sum_{\sigma\in\{\uparrow,\downarrow\}} \sum_{j,j'=1}^4\sum_{\ell,\ell'=1}^N
	h_{\sigma, j,j',\ell,\ell'}\hat{c}^\dagger_{\sigma, j,\ell}\hat{c}_{\sigma, j',\ell'} \;,
\end{eqnarray}
where $\sigma\in\{\uparrow,\downarrow\}$ is the spin index, $j\in\{1,2,3,4\}$ is the site index in the unit cell and $\ell$ is the index of the lattice translation $\tau_\ell=4a\ell$, where $a$ is the Mn-Mn distance and $\ell$ is an integer. 
We consider periodic boundary conditions with $N$ unit cells. 
The limit $N\rightarrow\infty$ is taken. 
The position of a Mn site is $R_{j,\ell}=aj+\tau_\ell$.

With $\hat{c}^{\dagger}_{\sigma,j,\ell}$ and $\hat{c}^{\phantom{\dagger}}_{\sigma,j,\ell}$, we denote the creation and annihilation operators of the Mn-$e_g$ orbital at $R_{j,\ell}$ pointing along the chain.

In a Bloch representation, the creation and annihilation operators $\hat b^\dagger_{\sigma,j}(k)$ and $\hat b_{\sigma,j}(k)$ are defined via
\begin{eqnarray}
	\label{equ:LBE-trafo-Bloch}
	\hat{b}^\dagger_{\sigma, j}(k) := 
	\sqrt{\frac{4a}{2\pi}}
	\sum_{\ell=1}^N e^{i\newk\tau_\ell} \hat{c}^\dagger_{\sigma, j,\ell} \, ,
\end{eqnarray}
and the $k$-points spacing is $\Delta_k=\frac{2\pi}{4aN}$. 
The $k$ points are chosen from the interval $k\in[-\frac{\pi}{4a},\frac{\pi}{4a}]$.

In this basis, the noninteracting Hamiltonian is
\begin{eqnarray}
	\hat{H}_0
	&=&\int dk\; \sum_{\sigma\in\{\uparrow,\downarrow\}}\sum_{j,j'=1}^4
	h_{\sigma, j,j'}(k)\hat{b}^\dagger_{\sigma,j}(k)\hat{b}_{\sigma,j'}(k),
\end{eqnarray}
with 
\begin{eqnarray}
	h_{\sigma, j,j'}(k)=\sum_{\ell=1}^N h_{\sigma,j,j',\ell,0}e^{ik\tau_\ell}\;.
\end{eqnarray}

The $k$-dependent Hamiltonian has the form
\begin{eqnarray}
	h_{\sigma}(k)=\left(\begin{array}{cccc}
	\sigma\Delta/2 & -t_{hop} & 0 & -t_{hop}\e{i4ak} \\
	-t_{hop}& \sigma\Delta/2 & -t_{hop} & 0  \\
	0& -t_{hop}& -\sigma\Delta/2 & -t_{hop}   \\
	 -t_{hop}\e{-i4ak} &0& -t_{hop}& -\sigma\Delta/2   \\
	\end{array}\right)\;,
	\nonumber\\
\end{eqnarray}
where we use $\sigma=1$ for the Hamiltonian of the spin-up electrons and $\sigma=-1$ for that of the spin down electrons. 

Diagonalization yields the eigenvalues $\epsilon_{\sigma,\nu}$ and the unitary matrix $T_{\sigma,j,\nu}$ holding the eigenvectors
\begin{eqnarray}
	\sum_{j'=1}^4 h_{\sigma,j,j'}(k) T_{\sigma,j',\nu}(k)=T_{\sigma,j,\nu}(k)\epsilon_{\sigma,\nu}(k) \, .
\end{eqnarray}
This leads to the band structure Eq.~\eqref{equ:1p-band-structure}.

The Hamilton operator is brought into the diagonal form using the creation and annihilation operators for specific bands
\begin{eqnarray}
	\hat{a}^\dagger_{\sigma,\nu}(k)=\sum_{j=1}^4 \hat{b}^\dagger_{\sigma,j}(k) T_{\sigma, j,\nu}(k)\;.
\end{eqnarray}

This yields
\begin{eqnarray}
	\hat{H}_0
	=\sum_\nu\sum_\sigma\int dk\; \epsilon_\nu(k) \hat{n}_{\sigma,\nu}(k)\;,
\end{eqnarray}
where $\hat{n}_{\sigma,\nu}(k):=\hat{a}^\dagger_{\sigma,\nu}(k)\hat{a}_{\sigma,\nu}(k)$ is the number operator for a particle in band $\nu$ and with wave vector $k$.

The interaction has the Hubbard form
\begin{eqnarray}
	\Hint &=& \sum_{\ell=1}^N\sum_{j=1}^4 U
	\hat{c}^\dagger_{\uparrow,j,\ell}
	\hat{c}^\dagger_{\downarrow,j,\ell}
	\hat{c}_{\downarrow,j,\ell}
	\hat{c}_{\uparrow,j,\ell}\;,
\end{eqnarray}
which yields
\begin{eqnarray}
	\Hint
	&=&
	\frac{4aU}{2\pi}
	\sum_{\nu_1,\nu_2,\nu_3,\nu_4}
	\int dk_1\ldots\int dk_4\;
	\nonumber\\
	&\times& \Phi_{\vec{\nu},\vec{k}}\sum_{n\in\mathds{Z}}\delta(P_{\vec{k}}+G_n)
	\nonumber\\
	&\times&
	\hat{a}^\dagger_{\uparrow,\nu_1}(k_1)
	\hat{a}^\dagger_{\downarrow,\nu_2}(k_2)
	\hat{a}_{\downarrow,\nu_3}(k_3)
	\hat{a}_{\uparrow,\nu_4}(k_4)\;,
\end{eqnarray}
with
\begin{eqnarray}
	\label{equ:LBE-Phi_k}
	\Phi_{\vec{\nu},\vec{k}}
	:=\sum_{j=1}^4T^*_{\uparrow, j,\nu_1}(k_1)T^*_{\downarrow, j,\nu_2}(k_2)
	T_{\downarrow, j,\nu_3}(k_3)T_{\uparrow, j,\nu_4}(k_4)
	\nonumber\\
\end{eqnarray}
and
\begin{eqnarray}
	P_{\vec{k}}:=k_1+k_2-k_3-k_4 \, .
\end{eqnarray}
With $G_n=\frac{2\pi}{4a}n$, we denote the reciprocal lattice vectors. 
The terms with nonzero reciprocal lattice vectors describe Umklapp processes, for which part of the momentum of the scattering particles is absorbed by the lattice.

Band structure and interaction determine the collision term $\Icoll$ of the Boltzmann equation, leading to Eq.~\eqref{equ: multi-band-BE 0}.
\section{Momentum distribution for $U=0$}
\label{app:U0}
\begin{figure}[]
	\includegraphics[width=0.48\textwidth]{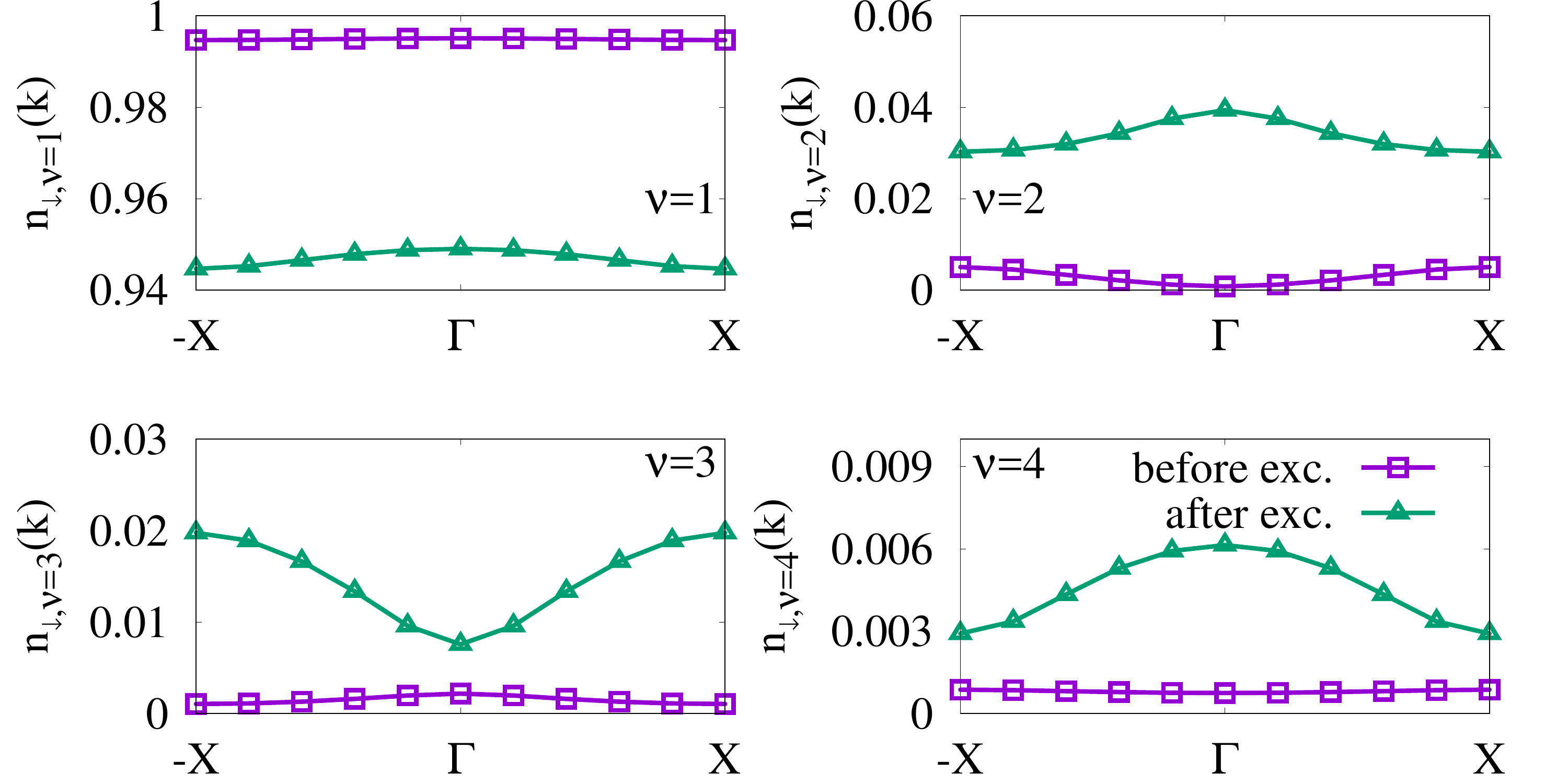}
	\caption
	{
		Momentum distribution for a system with open boundary conditions, $L=40$, $\Delta/\hopping = 2.3$, and $U/\hopping = 0$ before (magenta) and just after (green) the photoexcitation by applying operator Eq.~\eqref{eq:single_exciton} at the center of the system as obtained by the DMRG.
	}
	\label{fig:N_el_t0_Delta2.3_U0.0_OBC}
\end{figure}

\begin{figure}[]
	\includegraphics[width=0.48\textwidth]{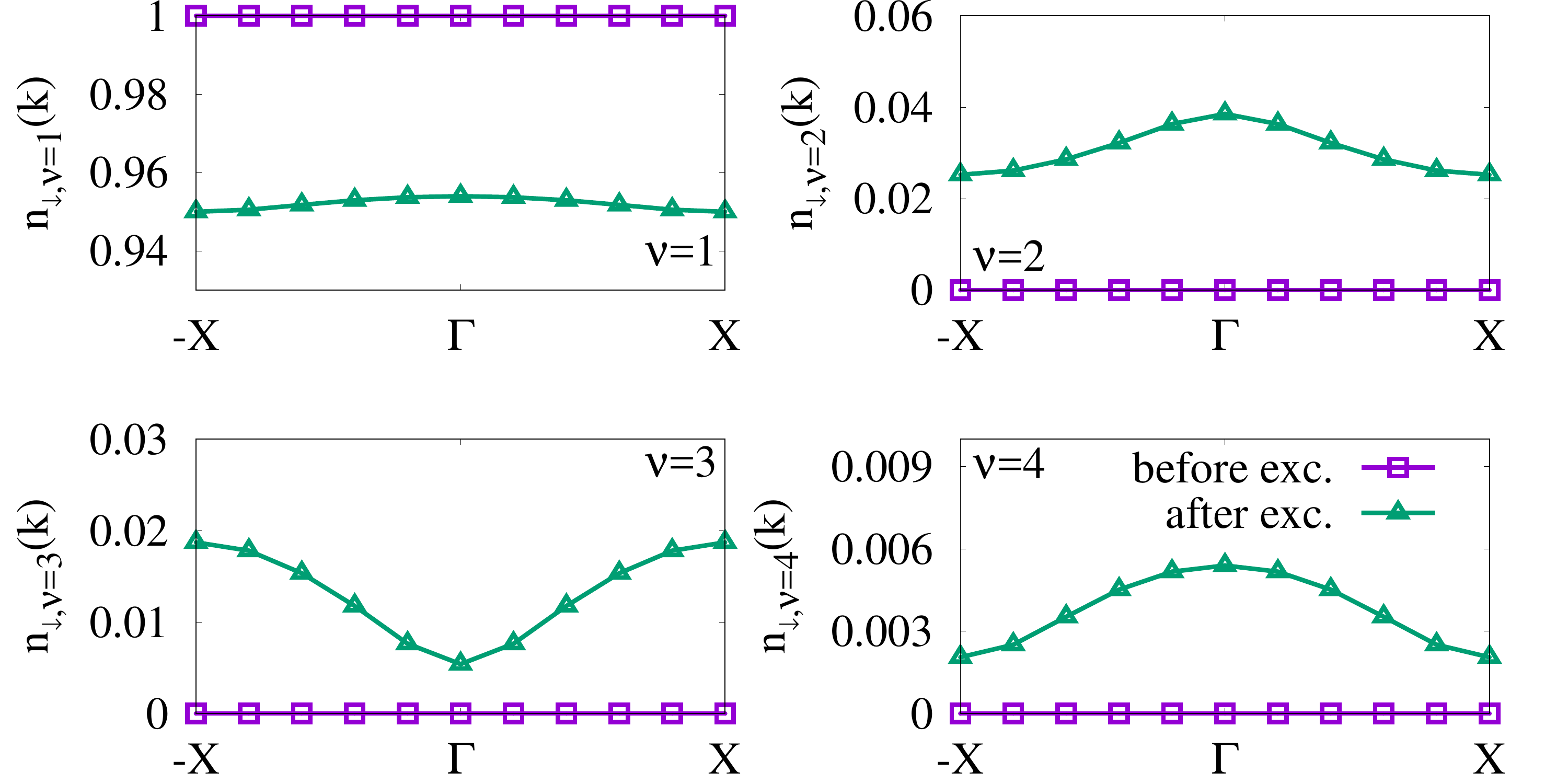}
	\caption
	{
		Momentum distribution as in \cref{fig:N_el_t0_Delta2.3_U0.0_OBC} but with periodic boundary conditions.
	}
	\label{fig:N_el_t0_Delta2.3_U0.0_PBC}
\end{figure}

In this appendix we show that the population inversion seen in Fig.~\ref{fig:N_el_t0_Delta2} at $U/\hopping = 4.3$ vanishes for $U=0$.
In \cref{fig:N_el_t0_Delta2.3_U0.0_OBC} the momentum distribution for a system similar to \cref{fig:N_el_t0_Delta2} with open boundary conditions, but with $U=0$, is shown. 
Without interactions, we expect at quarter filling that before the excitation the lowest band is completely filled and the other three bands completely empty. 
We obtain small differences to this expectation, which we associate with the choice of boundary conditions: In \cref{fig:N_el_t0_Delta2.3_U0.0_PBC} we present results for the same parameters, but with periodic boundary conditions.
As can be seen, here the expectation is perfectly matched. 
Note that the effect of the excitation is independent of the boundary conditions used. 
We see that for $\Delta / \hopping = 2.3$ at $U=0$ particles are excited from the lowest band to all higher bands. 
The resulting distributions show a peak at the $\Gamma$-point in the first, second, and fourth band, while in the third band a minimum is obtained.
The differences to this behavior visible in Fig.~\ref{fig:N_el_t0_Delta2} we associate to the effect of a finite value of $U/\hopping$.

\section{Estimating temperature and energy density of the excitation using MPS}
\label{appendix_entangler}
\label{finite_temp_app}
In this appendix we discuss how we obtained the values for the inverse temperature $\beta$ and the energy density due to the excitation, Eq.~\eqref{eq:single_exciton}. 

\subsection*{Final temperature of the excited state}
We use the purification approach discussed, e.g., in Refs.~\onlinecite{Schollwock:2011p2122} to compute the properties of the equilibrium state at finite temperatures.
In this approach the complete calculation takes place in an enlarged Hilbert space, which is created by adding an ancilla site to each physical site.
To obtain the state at a given temperature we perform an imaginary time evolution starting from a suitable state at infinite temperature ($\beta = 0$). 

Reference~\onlinecite{PhysRevB.93.045137} presents a way to obtain such a state, which conserves the two $U(1)$ symmetries of the model (total spin and particle number conservation) independently within the physical and the ancilla system.
The idea is to formulate a so-called entangler Hamiltonian or entangler, whose ground state is the desired state at $\beta=0$.
Note that the entangler is constructed only by fixing the particle statistics, i.e., $S=1/2$ fermions in our case.
Therefore the entangler can be used for any Hubbard like system to obtain an infinite temperature state.
We follow Ref.~\onlinecite{PhysRevB.93.045137} and use the entangler Hamiltonian
\begin{equation}
	\hat H_{\text{C2}}^{\text{Spin-}\frac12\text{-fermions}} = - \sum_{i\neq j,\, \sigma=\uparrow, \downarrow}\hat\Lambda^{\dagger}_{\sigma, i}\hat\Lambda^{\phantom{\dagger}}_{\sigma, j} + h.c.
	\label{eq:entanglerH}
\end{equation}
with $\hat \Lambda_{\sigma, i}=\hat c_{\sigma, i}\hat c_{\bar\sigma, a(i)} \hat P^\sigma_i$ and $\hat P^\sigma_i=\lvert 1 - \hat n_{\bar \sigma, i} - \hat n_{\sigma, a(i)}\lvert$.
The index $i$ labels a site in the physical space, the index $a(i)$ the corresponding site on the ancilla space.
In our MPS approach, we first formulate this long-range-interaction Hamiltonian as a finite state automaton (see Ref.~\onlinecite{SciPostPhys.3.5.035} for details).
In order to do so, we need to rewrite the projector, because it is not possible to evaluate the absolute value of a sum of operators within the finite state automaton framework.
This leads to 
\begin{eqnarray}
	\hat P^\sigma_i &=&\lvert 1 - \hat n_{\bar \sigma, i} - \hat n_{\sigma, a(i)}\lvert = (\hat P^\sigma_i)^2 \nonumber\\
	&=&1 - 2\hat n_{\bar \sigma, i} - 2\hat n_{\sigma, a(i)} +\hat n_{\bar \sigma, i}\hat n_{\sigma, a(i)} \nonumber\\
	&&+ \hat n_{\sigma, a(i)}\hat n_{\bar \sigma, i} + \hat n_{\bar \sigma, i}^2 + \hat n_{\sigma, a(i)}^2 \nonumber\\
	&=&1 - \hat n_{\bar \sigma, i} - \hat n_{\sigma, a(i)}+2\hat n_{\bar \sigma, i}\hat n_{\sigma, a(i)} \;.
\end{eqnarray}

\begin{figure}[h]
	\includegraphics[width=0.48\textwidth]{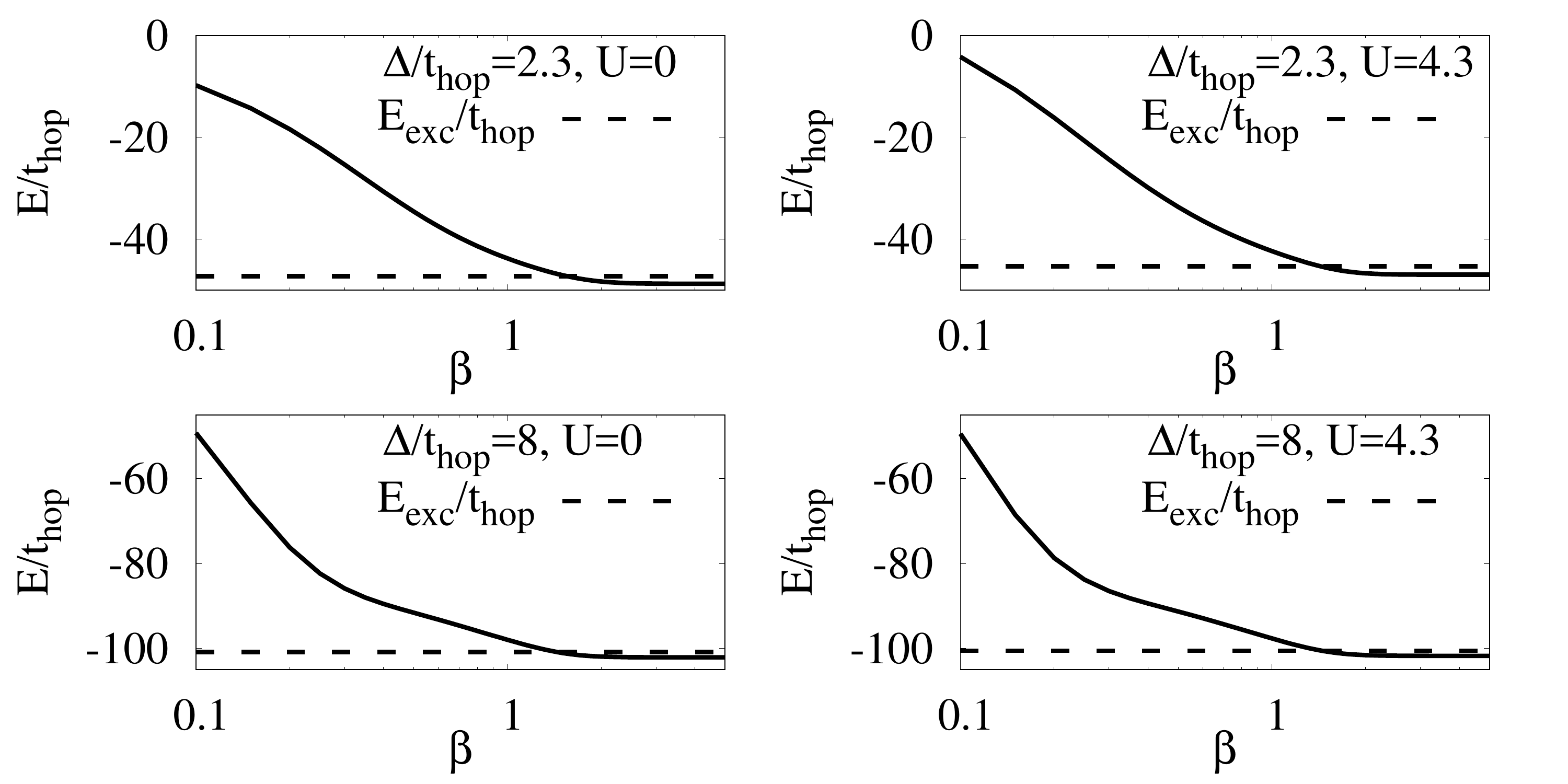}
	\caption
	{
		Energy of systems with $L=40$, $\Delta/\hopping=2.3, \, 8$, and $U/\hopping = 0, \, 8$ as a function of the inverse temperature $\beta$. 
		The results are obtained using DMRG by an imaginary time evolution starting from the ground state of \eqref{eq:entanglerH}, which is a suitable state with $\beta = 0$ in the physical space.
		The dashed horizontal lines indicate the energy after the excitation, which is obtained by computing the expectation value $E_{\rm exc} = \langle \psi(t) | \hat H | \psi(t) \rangle = const.$, with $\hat H$ the Hamiltonian \eqref{equ: eff model Hamiltonian} and $|\psi(t) \rangle$ the state after the excitation at time $t$. 
		The intersection of the finite-temperature results and $E_{\rm exc}$ indicates the value of $\beta$, which can be associated to the energy of the excitation. 
	}
	\label{fig:Energy_vs_beta}
\end{figure}

In \cref{fig:Energy_vs_beta} the imaginary time evolution starting from an infinite temperature state obtained as ground state of \eqref{eq:entanglerH} is shown for different values of $\Delta/\hopping$ and $U/\hopping$.

In thermal equilibrium, the value of $E(\beta)$ shown in \cref{fig:Energy_vs_beta} and of the energy of the excited state $E_{\rm exc} = \langle \psi(t) | \hat H | \psi(t) \rangle = const.$ is the same for the Hamiltonian \eqref{equ: eff model Hamiltonian}. 
Hence, the value of $\beta$ for which $E(\beta) = E_{\rm exc}$ corresponds to the temperature of the system after equilibration. 
The values for $E_{\rm exc}$ are displayed in \cref{tab:excited_energy}, the corresponding values of $\beta$ are shown in \cref{tab:excitedstatetemperature}.
\begin{table}[!htb]
	\caption
	{
		\label{tab:excited_energy}
		Values of the energy of the excited states.
	}
	\begin{tabular}{|l|cc|}
		\hline
		\hline
		$E_{\text{exc}}/\hopping$	&	$U/\hopping=0$	&	$U/\hopping=4.3$	\\
		\hline
		$\Delta/\hopping=2.3$		&	$-47.302$		&	$-45.354$	\\
		$\Delta/\hopping=8$			&	$-100.90$		&	$-100.55$	\\
		\hline
		\hline
	\end{tabular}
\end{table}
\begin{table}[!htb]
	\caption
	{
		\label{tab:excitedstatetemperature}
		Values of the inverse temperature $\beta$ at which $E(\beta) = E_{\rm exc}$.
	}
	\begin{tabular}{|l|cc|}
		\hline
		\hline
		$\beta$					&	$U/\hopping=0$	&	$U/\hopping=4.3$	\\
		\hline
		$\Delta/\hopping=2.3$	&	$1.56$			&	$1.43$	\\
		$\Delta/\hopping=8$		&	$1.45$			&	$1.44$	\\
		\hline
		\hline
	\end{tabular}
\end{table}

\subsection*{Estimating the energy density of the excitation}

We estimate the energy density of the excitation by considering the difference of the energy of the excited state to the ground state, $E_{\rm exc} - E_0$, and dividing it by the length of the finite system used in our simulations.
In this approach, we assume that the finite system considered represents a typical part of the lattice, which is excited by the incoming light, so that the energy density of the finite system would correspond to the one of an infinite lattice. 

Furthermore, we assume that the intensity $I$ of the incoming light amounts to the same energy density. 
We hence obtain
\begin{eqnarray}
	I = \frac{E_{\text{groundstate}}-E_{\text{excited state}}}{a\,L\,\tau}\;.
\end{eqnarray}

We use as value for the Mn-Mn distance $a=3.818$~\AA{}, see Ref.~\onlinecite{1985JMMM...53..153J}.
The duration of the light pulse is estimated to be $\tau=1$~fs.
The values for the ground state energies are given in \cref{tab:gs_energy} and are obtained via DMRG for chains with $L=40$ sites.

\begin{table}[!htb]
	\caption
	{
		\label{tab:gs_energy}
		Energy of the ground states.
	}
	\begin{tabular}{|l|cc|}
		\hline
		\hline
		$E_{\text{gs}}/\hopping$	&	$U/\hopping=0$	&	$U/\hopping=4.3$	\\
		\hline
		$\Delta/\hopping=2.3$		&	$-48.753$		&	$-46.973$	\\
		$\Delta/\hopping=8$			&	$-102.11$		&	$-101.76$	\\
		\hline
		\hline
	\end{tabular}
\end{table}

This leads to an intensity $\sim10^8$~W/mm. 
In a pump-probe setup, this would be the intensity of the pump laser in case of perfect absorption of the pump pulse. 
This value hence serves as a lower bound for the intensity needed to reproduce a scenario similar to the one discussed in this paper. 
As the intensity of lasers with ultrashort light pulses can reach $\sim10$ TW/mm, the estimate shows that similar investigations are within reach of typical pump-probe setups.

\end{appendix}

\bibliographystyle{prsty}
\bibliography{total}

\begin{thebibliography}{10}

\bibitem{Rini2007}
M. Rini, R. Tobey, N. Dean, J. Itatani, Y. Tomioka, Y. Tokura, R.~W.
  Schoenlein, and A. Cavalleri, Nature {\bf 449},  72  (2007).

\bibitem{1367-2630-18-9-093028}
I. Avigo, S. Thirupathaiah, M. Ligges, T. Wolf, J. Fink, and U. Bovensiepen,
  New Journal of Physics {\bf 18},  093028  (2016).

\bibitem{Schmitt1649}
F. Schmitt, P.~S. Kirchmann, U. Bovensiepen, R.~G. Moore, L. Rettig, M. Krenz,
  J.-H. Chu, N. Ru, L. Perfetti, D.~H. Lu, M. Wolf, I.~R. Fisher, and Z.-X.
  Shen, Science {\bf 321},  1649  (2008).

\bibitem{Tao62}
Z. Tao, C. Chen, T. Szilv{\'a}si, M. Keller, M. Mavrikakis, H. Kapteyn, and M.
  Murnane, Science {\bf 353},  62  (2016).

\bibitem{Mitrano2016}
M. Mitrano, A. Cantaluppi, D. Nicoletti, S. Kaiser, A. Perucchi, S. Lupi, P.
  Di~Pietro, D. Pontiroli, M. Ricc{\`o}, S.~R. Clark, D. Jaksch, and A.
  Cavalleri, Nature (London) {\bf 530},  461  (2016).

\bibitem{Hu2014}
W. Hu, S. Kaiser, D. Nicoletti, C.~R. Hunt, I. Gierz, M.~C. Hoffmann, M.
  Le~Tacon, T. Loew, B. Keimer, and A. Cavalleri, Nat Mater {\bf 13},  705
  (2014).

\bibitem{PhysRevLett.111.016401}
Z. Lenar\v{c}i\v{c} and P. Prelov\v{s}ek, Phys. Rev. Lett. {\bf 111},  016401
  (2013).

\bibitem{PhysRevB.92.201104}
Z. Lenar\ifmmode \check{c}\else \v{c}\fi{}i\ifmmode~\check{c}\else \v{c}\fi{},
  M. Eckstein, and P. Prelov\ifmmode~\check{s}\else \v{s}\fi{}ek, Phys. Rev. B
  {\bf 92},  201104  (2015).

\bibitem{Eckstein2013}
M. Eckstein and P. Werner, Physical review letters {\bf 110},  126401  (2013).

\bibitem{Eckstein2016}
P. Werner and M. Eckstein, Structural Dynamics {\bf 3},  023603  (2016).

\bibitem{Fausti189}
D. Fausti, R.~I. Tobey, N. Dean, S. Kaiser, A. Dienst, M.~C. Hoffmann, S. Pyon,
  T. Takayama, H. Takagi, and A. Cavalleri, Science {\bf 331},  189  (2011).

\bibitem{Zhao2005}
K. {Zhao}, Y.-H. {Huang}, H.-B. {L{\"u}}, M. {He}, K.-J. {Jin}, Z.-H. {Chen},
  Y.-L. {Zhou}, B.-L. {Cheng}, S.-Y. {Dai}, and G.-Z. {Yang}, Chinese Physics
  {\bf 14},  420  (2005).

\bibitem{Zhao2006}
K. {Zhao}, K.-j. {Jin}, H. {Lu}, Y. {Huang}, Q. {Zhou}, M. {He}, Z. {Chen}, Y.
  {Zhou}, and G. {Yang}, Applied Physics Letters {\bf 88},  141914  (2006).

\bibitem{Ni2012}
H. Ni, Z. Yue, K. Zhao, W. Xiang, S. Zhao, A. Wang, Y.-C. Kong, and H.-K. Wong,
  Opt. Express {\bf 20},  A406  (2012).

\bibitem{Snaith2013}
H.~J. Snaith, The Journal of Physical Chemistry Letters {\bf 4},  3623  (2013).

\bibitem{Gong2015}
J. Gong, S.~B. Darling, and F. You, Energy Environ. Sci. {\bf 8},  1953
  (2015).

\bibitem{Saucke2012}
G. Saucke, J. Norpoth, C. Jooss, D. Su, and Y. Zhu, Phys. Rev. B {\bf 85},
  165315  (2012).

\bibitem{Ifland2015}
B. Ifland, P. Peretzki, B. Kressdorf, P. Saring, A. Kelling, M. Seibt, and C.
  Jooss, Beilstein J. Nanotechnol. {\bf 6},  1467  (2015).

\bibitem{nanosecondlifetimes}
D. Raiser, S. Mildner, B. Ifland, M. Sotoudeh, P. Blöchl, S. Techert, and C.
  Jooss, Adv. Energy Mater. {\bf 7},  1602174  (2017).

\bibitem{shockley61_jap32_510}
W. Shockley and H. Queisser, J. Appl. Phys {\bf 32},  510  (1961).

\bibitem{white1992}
S.~R. White, Phys. Rev. Lett. {\bf 69},  2863  (1992).

\bibitem{white1993}
S.~R. White, Phys. Rev. B {\bf 48},  10345  (1993).

\bibitem{Schollwock:2005p2117}
U. Schollw{\"o}ck, Rev. Mod. Phys. {\bf 77},  259  (2005).

\bibitem{daley04}
A.~J. Daley, C. Kollath, U. Schollw\"{o}ck, and G. Vidal, J. Stat. Mech.:
  Theor. Exp. {\bf 2004},  P04005  .

\bibitem{white04}
S.~R. White and A.~E. Feiguin, Phys. Rev. Lett. {\bf 93},  076401  (2004).

\bibitem{Schollwock:2011p2122}
U. Schollw{\"o}ck, Annals of Physics {\bf 326},  96  (2011).

\bibitem{Biebl2016}
F.~R.~A. Biebl and S. Kehrein, Phys. Rev. B {\bf 95},  104304  (2017).

\bibitem{sotoudeh17_prb95_235150}
M. Sotoudeh, S. Rajpurohit, P. Bl\"ochl, D. Mierwaldt, J. Norpoth, V. Roddatis,
  S. Mildner, B. Kressdorf, B. Ifland, and C. Jooss, Phys. Rev. B {\bf 95},
  235150  (2017).

\bibitem{Hotta2004}
T. Hotta and E. Dagotto,  in {\em Colossal Magnetoresistive Manganites}, edited
  by T. Chatterji (Springer Netherlands, Dordrecht, 2004), pp.\ 207--262.

\bibitem{hansgerd1}
D.~R. Neuber, M. Daghofer, H.~G. Evertz, W. von~der Linden, and R.~M. Noack,
  Phys. Rev. B {\bf 73},  014401  (2006).

\bibitem{Dagotto2003}
E. Dagotto, {\em Nanoscale Phase Separation and Colossal Magnetoresistance}
  (Springer, Berlin, Heidelberg, New York, 2003).

\bibitem{hohenberg64_pr136_864}
P. Hohenberg and W. Kohn, Phys. Rev. {\bf 136},  B864  (1964).

\bibitem{kohn65_pr140_A1133}
W. Kohn and L.~J. Sham, Phys. Rev. {\bf 140},  A1133  (1965).

\bibitem{becke93_jcp98_1372}
A.~D. Becke, J. Chem. Phys. {\bf 98},  1372  (1993).

\bibitem{anisimov91_prb44_943}
V.~I. Anisimov, J. Zaanen, and O.~K. Andersen, Phys. Rev. B {\bf 44},  943
  (1991).

\bibitem{hedin65_pr139_796}
L. Hedin, Phys. Rev. {\bf 139},  A796  (1965).

\bibitem{Kanamori1963}
J. Kanamori, Progress of Theoretical Physics {\bf 30},  275  (1963).

\bibitem{bloechl94_prb50_17953}
P.~E. Bl\"ochl, Phys. Rev. B {\bf 50},  17953  (1994).

\bibitem{car85_prl55_2471}
R. Car and M. Parrinello, Phys. Rev. Lett. {\bf 55},  2471  (1985).

\bibitem{Note1}
We have done calculations for different unit cells. We choose a twelve-site
  model because it produces only a few frustrated structures ($N_s$ must be
  divisible by $2$, $3$, and $4$.) The size of the unit cell has been chosen
  small to avoid being trapped in metastable states.

\bibitem{zener51_pr82_403}
C. Zener, Phys. Rev. {\bf 82},  403  (1951).

\bibitem{zhou00_prb62_3834}
J.-S. Zhou and J.~B. Goodenough, Phys. Rev. B {\bf 62},  3834  (2000).

\bibitem{Silva2010}
L.~G. G.~V. {Dias da Silva}, K.~A. Al-Hassanieh, A.~E. Feiguin, F.~A. Reboredo,
  and E. Dagotto, Physical Review B {\bf 81},  125113  (2010).

\bibitem{PhysRevB.85.205127}
F. Hofmann and M. Potthoff, Phys. Rev. B {\bf 85},  205127  (2012).

\bibitem{Wall2011}
S. Wall, D. Brida, S.~R. Clark, H.~P. Ehrke, D. Jaksch, A. Ardavan, S. Bonora,
  H. Uemura, Y. Takahashi, T. Hasegawa, {\it et~al.}, Nature Physics {\bf 7},
  114  (2011).

\bibitem{Dagotto2008}
K.~A. Al-Hassanieh, F.~A. Reboredo, A.~E. Feiguin, I. Gonz{\'{a}}lez, and E.
  Dagotto, Phys. Rev. Lett. {\bf 100},  1  (2008).

\bibitem{Zaletel2015}
M.~P. Zaletel, R.~S. Mong, C. Karrasch, J.~E. Moore, and F. Pollmann, Physical
  Review B {\bf 91},  165112  (2015).

\bibitem{PhysRevA.78.012356}
G.~M. Crosswhite and D. Bacon, Phys. Rev. A {\bf 78},  012356  (2008).

\bibitem{SciPostPhys.3.5.035}
S. Paeckel, T. Köhler, and S.~R. Manmana, SciPost Phys. {\bf 3},  035  (2017).

\bibitem{Amico:2008en}
L. Amico, R. Fazio, A. Osterloh, and V. Vedral, Rev. Mod. Phys. {\bf 80},  517
  (2008).

\bibitem{scipal}
S.~C. Kramer, {\em CUDA-based Scientific Computing: Tools and Selected
  Applications} (Nieders{\"a}chsische Staats- und Universit{\"a}tsbibliothek,
  G{\"o}ttingen, 2013).

\bibitem{Note2}
{c}uBLAS is an implementation of the BLAS library especially for the usage on
  NVIDIA CUDA devices.

\bibitem{Bedurftig:1998p457}
G. Bed{\"u}rftig, B. Brendel, H. Frahm, and R.~M. Noack, Phys. Rev. B {\bf 58},
   10225  (1998).

\bibitem{White:2002p348}
S.~R. White, I. Affleck, and D.~J. Scalapino, Phys. Rev. B {\bf 65},  165122
  (2002).

\bibitem{giamarchi}
T. Giamarchi, {\em Quantum Physics in One Dimension}, Vol.~121 of {\em
  International Series of Monographs on Physics} (Oxford University Press,
  Oxford, 2004).

\bibitem{HubbBook}
F.~H.~L. Essler, H. Frahm, F. G{\"o}hmann, A. Kl{\"u}mper, and V.~E. Korepin,
  {\em The One-Dimensional {H}ubbard Model} (Cambridge University Press,
  Cambridge, 2005).

\bibitem{LiebRobinson72}
E.~H. Lieb and D.~W. Robinson, Commun. Math. Phys. {\bf 28},  251  (1972).

\bibitem{preprintLorenzo}
L. {Cevolani}, J. {Despres}, G. {Carleo}, L. {Tagliacozzo}, and L.
  {Sanchez-Palencia}, arXiv:1706.00838  (2017).

\bibitem{RevModPhys.75.473}
A. Damascelli, Z. Hussain, and Z.-X. Shen, Rev. Mod. Phys. {\bf 75},  473
  (2003).

\bibitem{Spohn2012}
M.~L.~R. F\"urst, C.~B. Mendl, and H. Spohn, Phys. Rev. E {\bf 86},  031122
  (2012).

\bibitem{Spohn2013}
M.~L.~R. F\"urst, C.~B. Mendl, and H. Spohn, Phys. Rev. E {\bf 88},  012108
  (2013).

\bibitem{Erdoes2004}
L. Erd\"os, M. Salmhofer, and H.-T. Yau, Journal of Statistical Physics {\bf
  116},  367  (2004).

\bibitem{Spohn2009}
J. Lukkarinen and H. Spohn, Journal of Statistical Physics {\bf 134},  1133
  (2009).

\bibitem{Kamenev2009}
A. Kamenev and A. Levchenko, Advances in Physics {\bf 58},  197  (2009).

\bibitem{Essler2014}
F.~H.~L. Essler, S. Kehrein, S.~R. Manmana, and N.~J. Robinson, Phys. Rev. B
  {\bf 89},  165104  (2014).

\bibitem{Haug1996}
H. Haug and A.-P. Jauho, {\em Quantum Kinetics in Transport and Optics of
  Semiconductors} (Springer, Berlin, Heidelberg, 1996).

\bibitem{PhysRevB.93.045137}
A. Nocera and G. Alvarez, Phys. Rev. B {\bf 93},  045137  (2016).

\bibitem{1985JMMM...53..153J}
Z. {Jir{\'a}k}, S. {Krupi{\v c}ka}, Z. {{\v S}im{\v s}a}, M. {Dlouh{\'a}}, and
  S. {Vratislav}, Journal of Magnetism and Magnetic Materials {\bf 53},  153
  (1985).

\end{thebibliography}

\end{document}